\documentclass[11pt,a4paper]{amsart}
\usepackage{mathtools}
\usepackage{dsfont,amsfonts}
\usepackage[left=25mm,top=25mm,bottom=25mm,right=25mm]{geometry}
\usepackage[english]{babel} 
\usepackage[utf8]{inputenc} 
\usepackage{cmap}           
\usepackage{hyperxmp}       
\usepackage[pdfdisplaydoctitle=true,
colorlinks=true,
urlcolor=blue,
citecolor=blue,
linkcolor=blue,
pdfstartview=FitH,
pdfpagemode=UseNone,
bookmarksnumbered=true]{hyperref} 
\usepackage{booktabs}

\newtheorem{thm}{Theorem}[section]

\theoremstyle{definition}

\newtheorem{rem}[thm]{Remark}
\newtheorem{exa}[thm]{Example}

\newcommand{\RR}{\mathbb{R}}                
\newcommand{\ZZ}{\mathbb{Z}}                

\newcommand{\Fix}{\mathbb{F}\mathrm{ix}}    
\newcommand{\HH}{\mathbb{H}}                
\newcommand{\Sym}{\mathbb{S}}               

\newcommand{\OO}{\mathrm{O}}                
\newcommand{\SO}{\mathrm{SO}}               
\newcommand{\octa}{\mathbb{O}}              
\newcommand{\tetra}{\mathbb{T}}             
\newcommand{\DD}{\mathbb{D}}                
\newcommand{\triv}{\mathds{1}}		        
\newcommand{\id}{e}                	        

\newcommand{\strata}[1]{\Sigma_{[#1]}}	                

\newcommand{\ee}{\pmb{e}}                   
\newcommand{\nn}{\pmb{n}}                   
\newcommand{\vv}{\pmb{v}}                   

\newcommand{\ba}{\mathbf{a}}
\newcommand{\bb}{\mathbf{b}}
\newcommand{\bc}{\mathbf{c}}
\newcommand{\bd}{\mathbf{d}}

\newcommand{\Id}{\mathbf{1}}                
\newcommand{\bh}{\mathbf{h}}
\newcommand{\bk}{\mathbf{k}}

\newcommand{\bt}{\mathbf{t}}
\newcommand{\bv}{\mathbf{v}}

\newcommand{\bE}{\mathbf{E}}                
\newcommand{\bH}{\mathbf{H}}                
\newcommand{\bI}{\mathbf{I}}                
\newcommand{\bJ}{\mathbf{J}}
\newcommand{\bA}{\mathbf{A}}                
\newcommand{\bB}{\mathbf{B}}                %
\newcommand{\bS}{\mathbf{S}}                
\newcommand{\bT}{\mathbf{T}}                
\newcommand{\bC}{\mathbf{C}}                
\newcommand{\bP}{\mathbf{P}}                
\newcommand{\bR}{\mathbf{R}}                
\newcommand{\lc}{\pmb \varepsilon}          

\DeclareMathOperator{\tr}{tr}
\DeclareMathOperator{\3dots}{\raisebox{-0.25ex}{\vdots}}
\newcommand{\otimesbar}{\mathbin{\underline{\overline{\otimes}}}}
\newcommand{\rot}{\mathbf{r}}

\newcommand{\norm}[1]{\lVert#1\rVert}       
\newcommand{\abs}[1]{\lvert#1\rvert}        
\newcommand{\set}[1]{\left\{#1\right\}}     

\hypersetup{
    pdftitle={Upper bounds estimates of the distance to cubic or orthotropic elasticity},
    pdfauthor={Rodrigue Desmorat and Boris Kolev},
    pdfsubject={},
    pdfkeywords={Symmetry class; Elasticity; Distance to a symmetry class; Cubic symmetry; Orthotropy; Upper bounds},
    pdflang=en
}

\begin{document}

\title[Upper bounds estimates of the distance to cubic or orthotropic elasticity]{Upper bounds estimates of the distance to cubic or orthotropic elasticity}%

\author{R. Desmorat}
\address[Rodrigue Desmorat]{Université Paris-Saclay, CentraleSupélec, ENS Paris-Saclay, CNRS, Laboratoire de Mécanique Paris-Saclay, 91190, Gif-sur-Yvette, France.}
\email{rodrigue.desmorat@ens-paris-saclay.fr}

\author{B. Kolev}
\address[Boris Kolev]{Université Paris-Saclay, CentraleSupélec, ENS Paris-Saclay, CNRS, Laboratoire de Mécanique Paris-Saclay, 91190, Gif-sur-Yvette, France.}
\email{boris.kolev@ens-paris-saclay.fr}

\date{June 3, 2022}%

\thanks{The authors were partially supported by CNRS Projet 80--Prime GAMM (Géométrie algébrique complexe/réelle et mécanique des matériaux).}


\begin{abstract}
  We address the problem, not of the determination--which usually needs numerical methods--but of an accurate analytical estimation of the distance of a raw elasticity tensor to cubic symmetry and to orthotropy. We point out that there are not one but several second-order tensors that carry the likely cubic/orthotropic coordinate system of the raw tensor. Since all the second-order covariants of an (exactly) cubic elasticity tensor are isotropic, distance estimates based only on such covariants are not always accurate. We extend to cubic symmetry and to orthotropy the technique recently suggested by Klimeš for transverse isotropy: solving analytically an auxiliary quadratic minimization problem whose solution is a second-order tensor that carries the likely cubic coordinate system. Numerical examples are provided, on which we evaluate the accuracy of different upper bounds estimates of the distance to cubic or orthotropic symmetry.
\end{abstract}

\maketitle

\section{Introduction}
\label{sec:intro}

The computation of the distance from a raw (measured) elasticity tensor $\bE_{0}$ to a chosen symmetry class is not an easy task when $\bE_{0}$ is expressed in an arbitrary coordinate system~\cite{Fra1995,DVS1998,FGB1998,Del2005,KS2008,KS2009,WGB2019,ADKP2022}. This problem raises questions, such as $(i)$ the proper definition of the distance function and $(ii)$ the choice of the related norm. Concerning the first point, the axioms defining a true distance between tensors~\cite{MN2006} are rarely recalled so that when non classical choices are made, the name \emph{distance} is sometimes used abusively or without proof that the formulated concept is indeed a distance (such as in~\cite{SMB2020}). Concerning the second point, and as pointed out in~\cite{MN2006,MGD2019},
many different norms can be used to define a distance to an elasticity symmetry class. In the present work, we will mainly consider the Euclidean norm and its related distance. We will address the case of the Log-Euclidean norm only in the Appendix.

It has been shown by Stahn and coworkers~\cite{SMB2020} that when analytical solutions of the distance problem  were available (an algorithm that only estimates this distance in their case), the engineering problem of the assignation of a symmetry class to a measured stiffness was simplified. Elasticity symmetry classes for which an analytical solution of the standard distance problem is available are scarce: distance to isotropy (in 2D and 3D), distance to plane elasticity square symmetry~\cite{Via1997} and distance to plane orthotropy~\cite{ADKD2021}. In 3D elasticity, it seems rather difficult to get an analytical solution of the standard distance to a symmetry class problem~\cite{FGB1998} (see~\cite{ADKP2022} for an attempt for cubic symmetry). This is the reason why the subject that we address here is the determination of accurate upper bounds estimates of the distance to a symmetry class.

Such upper bounds estimates have been proposed for 2D elasticity in~\cite{Via1997} and in \cite{OLDK2021} for orthotropy. For 3D elasticity, such bounds have been formulated in~\cite{Kli2018} for transverse isotropy, and in~\cite{SMB2020} for all symmetry classes, using a second-order harmonic component of the elasticity tensor introduced in~\cite{Bac1970} (in fact, a second-order covariant of $\bE_{0}$~\cite{OKDD2021}). This covariant is assumed to carry the likely symmetry coordinate system of $\bE_{0}$. Note however that all second-order covariants of an exactly cubic elasticity tensor are all isotropic. Therefore, for a material expected to be cubic (by its microstructure for instance), a methodology based on second-order covariants is probably meaningless.

The present work focuses on cubic symmetry, first. Our goal is to evaluate and improve the accuracy of upper bounds estimates of the distance to cubic symmetry, on three raw elasticity tensors, for various materials:
\begin{itemize}
  \item on the elasticity tensor measured by François and coworkers~\cite{FGB1998} for a Ni-based single crystal superalloy (such as CMSX-4), and expected to be cubic,
  \item on an academic elasticity tensor studied by Stahn and coworkers~\cite{SMB2020} (close to be orthotropic),
  \item on an elasticity tensor identified by François~\cite{Fra1995}, using ultrasonic measurements by Arts~\cite{Art1993}, for a Vosges sandstone.
\end{itemize}
These improvements are the following:
\begin{enumerate}
  \item the use of several second-order covariants of the raw elasticity tensor $\bE_{0}$, rather than one;
  \item the introduction of an auxiliary problem (a quadratic minimization problem which reduces to an eigenvalue problem) whose solution is a second-order symmetric tensor which carries the likely cubic coordinate system sought.
\end{enumerate}
Using a theoretical tool from~\cite{ADD2020}, the second improvement can be seen as an extension of a technique introduced recently by Klimeš~\cite{Kli2018} for transverse isotropy, to cubic and orthotropic symmetry.

We will finally show that, in practice, our list of second-order tensors, intended to carry the likely cubic coordinate system of $\bE_{0}$, produces also more accurate upper bounds estimates of the distance to 3D elasticity orthotropy.

\section{Geometry of the elasticity tensor}
\label{sec:InvCovEla}

In this paper, we use the notation $\Sym^{n}$ for the vector space of totally symmetric tensors of order $n$ and $\HH^{n}$ for its subspace of harmonic tensors (traceless totally symmetric tensors). Since, furthermore, all basis are assumed orthonormal, we shall not distinguish between covariant, contravariant or mixed tensors.

Let $\vv$ be a vector, $\ba$, $\bb$ or $\bc$ be a symmetric second-order tensor, $\bE$ be an elasticity tensor. The action of a rotation $r\in \SO(3)$ on these tensors, denoted by $\star$, is given by
\begin{equation}\label{eq:gE}
  (r\star \vv)_{i} := r_{ij}v_{j},
  \qquad
  (r\star \ba)_{ij} := r_{ik}r_{jl}a_{kl},
  \qquad
  (r\star \bE)_{ijkl} := r_{ip}r_{jq}r_{kr}r_{ls} E_{ijkl}.
\end{equation}

\subsection{Harmonic decomposition}

The second-order \emph{dilatation tensor} is defined as
\begin{equation*}
  \bd := \tr_{12}\bE \qquad \left(\emph{i.e.}, \; d_{ij} = E_{kkij}\right),
\end{equation*}
and the second-order \emph{Voigt tensor}, as
\begin{equation*}
  \bv := \tr_{13}\bE \qquad \left(\emph{i.e.}, \; v_{ij}=E_{kikj}\right).
\end{equation*}
The harmonic decomposition of $\bE$~\cite{Bac1970,Cow1989,Bae1993} corresponds to its splitting into
  \begin{equation}\label{eq:dec-harm-E}
    \bE = (\lambda, \mu, \bd^{\prime}, \bv^{\prime}, \bH),
  \end{equation}
  where
  \begin{equation*}
    \lambda = \frac{1}{15}(2 \tr\bd-\tr\bv)
    ,\qquad
    \mu = \frac{1}{30}(3\tr\bv- \tr\bd),
  \end{equation*}
  are two scalar invariants,
  \begin{equation*}
    \bd^{\prime} = \bd - \frac{1}{3} (\tr \bd) \, \Id, \qquad \bv^{\prime} = \bv - \frac{1}{3} (\tr \bv) \, \Id,
  \end{equation*}
  are two deviatoric tensors, and
  \begin{equation}\label{eq:H}
    \bH=\bE^{s}-(2\mu+ \lambda)\, \Id \odot \Id -
    \frac{2}{7}\Id \odot ( \bd^{\prime} +2  \bv^{\prime}),
  \end{equation}
  is an harmonic fourth-order tensor (\emph{i.e.}, totally symmetric and traceless). Here, $\bT^{s}$ means the total symmetrization of the fourth-order tensor $\bT$, and $\ba\odot\bb$ is the symmetrized tensor product of two second-order tensors, defined by
  \begin{equation*}
    \ba \odot \bb =(\ba \otimes \bb)^{s}.
  \end{equation*}

The explicit harmonic decomposition of $\bE$ is then
\begin{equation}\label{eq:DecompHarm}
  \bE = 2 \mu\, \bI + \lambda\, \Id \otimes \Id +\frac{2}{7}\, \Id \odot (\bd' +2 \bv' )
  +2\, \Id \otimes_{(2,2)} (\bd' - \bv' )+\bH ,
\end{equation}
which can also be written as,
\begin{align*}
  \bE = & 2 \mu\, \bI + \lambda\, \Id \otimes \Id +\bH
  \\
        & +
  \frac{1}{7}\left(\Id\otimes (5 \bd'-4 \bv')+ (5 \bd'-4 \bv') \otimes \Id
  +2 \,\Id \otimesbar (6  \bv' - 4 \bd')+2   (6  \bv' - 4 \bd') \otimesbar \Id \right),
\end{align*}
where, given $\ba$ and $\bb$ be two symmetric second-order tensors, $\otimes_{(2,2)}$ is the Young-symmetrized tensor product defined by
\begin{equation*}
  \ba \otimes_{(2,2)}\!\bb = \frac{1}{3} \big( \ba \otimes \bb +  \bb \otimes \ba
  - \ba \otimesbar \bb - \bb \otimesbar \ba \big),
\end{equation*}
with
\begin{equation*}
  (\ba \otimesbar \bb)_{ijkl} : = \frac{1}{2} (a_{ik}b_{jl} + a_{il}b_{jk}),
  \qquad
  I_{ijkl}=(\Id \otimesbar \Id)_{ijkl} = \frac{1}{2} (\delta_{ik}\delta_{jl} + \delta_{il}\delta_{jk}).
\end{equation*}

\begin{rem}\label{rem:normE2}
  When the Euclidean norm
  \begin{equation*}
    \norm{\bE}=\sqrt{\bE::\bE}=\sqrt{E_{ijkl}E_{ijkl}},
  \end{equation*}
  is used, one gets
  \begin{equation*}
    \norm{\bE}^{2} =3 \left(3 \lambda ^{2}+4 \lambda  \mu +8 \mu ^{2}\right) +  \frac{2}{21} \norm{\bd^{\prime}+2 \bv^{\prime}}^{2} +\frac{4}{3}\norm{\bd^{\prime}- \bv^{\prime}}^{2}+ \norm{\bH}^{2}.
  \end{equation*}
\end{rem}

\subsection{Symmetry groups, symmetry classes, symmetry strata}
\label{subsec:SymClasses}

Given an elasticity tensor $\bE$, its \emph{symmetry group} $G_{\bE}$ is defined as the set of all rotations $g$ such that
\begin{equation*}
  g\star \bE = \bE.
\end{equation*}
Given a second elasticity tensors $\overline{\bE}$, deduced from $\bE$ by a rotation $r$, \emph{i.e.}, $\overline{\bE} = r \star \bE$, then its symmetry group $G_{\overline{\bE}}$ is \emph{conjugate} to $G$, meaning that $G_{\overline{\bE}}=r G_{\bE} r^{-1}$, since
\begin{equation*}
  g \star \bE = \bE  \iff  (r g \,r^{-1})\star \overline{\bE} = \overline{\bE}.
\end{equation*}
Therefore, it is not really the symmetry group $G_{\bE}$ of $\bE$ which is meaningful by itself but its \emph{conjugacy class} $[G_{\bE}]$, which is the set of all subgroups which are conjugate to $G_{\bE}$. Such a set is called a \emph{symmetry class}. For each symmetry class, it is therefore useful to provide an explicit \emph{representative subgroup} $G$ in this class, which allows to visualize the corresponding symmetry (see \autoref{sec:groups}). Therefore, the set of all elasticity tensors $\bE$ whose symmetry group $G_{\bE}$ is conjugate to $G$, will be denoted by $\strata{G}$ and called the \emph{symmetry stratum} associated to the symmetry class $[G]$.

\begin{exa}
  The \emph{orthotropic symmetry class} has for \emph{canonical representative} the dihedral group
    \begin{equation}\label{eq:D2}
      \DD_{2} =  \set{\id, \rot(\ee_{1},\pi), \rot(\ee_{2},\pi), \rot(\ee_{3},\pi)},
    \end{equation}
 where $\rot(\nn,\theta)$ is the rotation of angle $\theta$ around $\nn$. The orthotropic stratum is thus denoted by $\strata{\DD_{2}}$.
\end{exa}

For elasticity tensors, Forte and Vianello~\cite{FV1996} have established that there are exactly eight symmetry classes: the triclinic class $[\triv]$, the monoclinic class $[\ZZ_{2}]$, the orthotropic class $[ \DD_{2}]$, the trigonal class $[\DD_{3}]$, the tetragonal class $[\DD_{4}]$, the cubic class $[\octa]$, the transverse isotropic class $[\OO(2)]$ and the isotropic class $[\SO(3)]$.

\begin{rem}
  The symmetry classes, their number, and their partial ordering are strongly dependent on the tensor type. For the natural action introduced in~\eqref{eq:gE}, we get the following.
  \begin{itemize}
    \item There are two symmetry classes for a vector $\vv$: $[\SO(2)]$ (axial symmetry, if $\vv \ne 0$) and $[\SO(3)]$ (isotropy, if $\vv=0$). Note that, since the transverse isotropy conjugacy class $[\OO(2)]$ is not a symmetry class in this case, a vector $\vv$ which is invariant by $\OO(2)$ (or a conjugate of $\OO(2)$) is thus isotropic and therefore vanishes, since the immediate symmetry class which contains $[\OO(2)]$ is $[\SO(3)]$;

    \item There are three symmetry classes for a symmetric second-order tensor $\ba$ (and for a deviatoric tensor $\ba'$): $[\DD_{2}]$ (orthotropy, if $\ba$ has three distinct eigenvalues), $[\OO(2)]$ (transverse isotropy, if $\ba$ has two distinct eigenvalues), and $[\SO(3)]$ (isotropy, if $\ba'=0$);

    \item The symmetry classes for an harmonic (totally symmetric and traceless) fourth-order tensor $\bH$ are the same eight symmetry classes as those of an elasticity tensor~\cite{IG1984,FV1996}: $[\triv]$, $[\ZZ_{2}]$, $[\DD_{2}]$, $[\DD_{3}]$, $[ \DD_{4}]$, $[\octa]$, $[\OO(2)]$ and $[\SO(3)]$ (isotropy, $\bH=0$).
  \end{itemize}
\end{rem}

An elasticity tensor $\bE$ in the symmetry stratum $\strata{G}$ may have exactly as symmetry group the \emph{canonical representative group} $G$ itself, and not a conjugate of $G$, that is
\begin{equation*}
  g\star \bE =\bE , \quad \forall g\in G.
\end{equation*}
\emph{i.e.}, $\bE$ is fixed by $G$. In that case, we say that $\bE$ is in its \emph{normal form} (or natural basis).
When $G=\octa$, elasticity tensors in cubic normal form are written as
\begin{equation}\label{eq:VoigtCubic}
  [\bE]=\begin{pmatrix}
    E_{1111} & E_{1122} & E_{1122} & 0          & 0          & 0          \\
    E_{1122} & E_{1111} & E_{1122} & 0          & 0          & 0          \\
    E_{1122} & E_{1122} & E_{1111} & 0          & 0          & 0          \\
    0        & 0        & 0        & 2 E_{1212} & 0          & 0          \\
    0        & 0        & 0        & 0          & 2 E_{1212} & 0          \\
    0        & 0        & 0        & 0          & 0          & 2 E_{1212}
  \end{pmatrix}.
\end{equation}
When $G=\DD_{2}$, elasticity tensors in orthotropic normal form are written as
\begin{equation}\label{eq:VoigtOrtho}
  [\bE]=\begin{pmatrix}
    E_{1111} & E_{1122} & E_{1133} & 0          & 0          & 0          \\
    E_{1122} & E_{2222} & E_{2233} & 0          & 0          & 0          \\
    E_{1133} & E_{2233} & E_{3333} & 0          & 0          & 0          \\
    0        & 0        & 0        & 2 E_{2323} & 0          & 0          \\
    0        & 0        & 0        & 0          & 2 E_{1313} & 0          \\
    0        & 0        & 0        & 0          & 0          & 2 E_{1212}
  \end{pmatrix}.
\end{equation}
Here, both normal forms are expressed in Kelvin matrix representation. Note that for each tensor $\overline{\bE} \in \strata{G}$, one can find a rotation $r$ such that $r \star \overline{\bE}$ is in its normal form.

The set of tensors $\bE$ which are fixed by a group $G$, not necessarily a representative symmetry group, is called the \emph{fixed point set} of $G$ and denoted by $\Fix(G)$. It is a linear subspace of the vector space of elasticity tensors. The orthogonal projection on this set will be denoted by $\bR_{G}$. It is uniquely defined, but can be expressed in several ways.
When the group $G$ is finite, this orthogonal projection can be recast as the averaging
\begin{equation}\label{eq:ReyG}
  \bR_{G}\left(\bE\right) = \frac{1}{\abs{G}}\sum_{g \in G} g \star \bE,
\end{equation}
where $\abs{G}$ is the cardinal of $G$, in which case, $\bR_{G}$ is called the \emph{Reynolds operator} associated with the group $G$ \cite[Chapter 2]{Stu2008}. Alternative expressions for this orthogonal projection onto $\Fix(G)$ are provided in \autoref{sec:Klimes} for transverse isotropy ($G$ conjugate to $\OO(2)$), in \autoref{sec:Upper bound-Cubic} for cubic symmetry ($G$ conjugate to $\octa$, remark \ref{rem:ReynolodsCubic}), and in \autoref{subsec:Reynolds-projections} for orthotropy ($G$ conjugate to $\DD_{2}$).

\begin{rem} \label{rem:RGEharm}
  Let $\bE=(\lambda, \mu, \bd^{\prime}, \bv^{\prime}, \bH)$ be the harmonic decomposition of an elasticity tensor $\bE$ and let $G$ be a subgroup of $\SO(3)$. The orthogonal projection on the vector space $\Fix(G)$, of tensors fixed by $G$, can be expressed as
  \begin{equation*}
    \bR_{G}(\bE) = \bR_{G}\left((\lambda, \mu, \bd^{\prime}, \bv^{\prime}, \bH)\right)= (\lambda, \mu, \bR_{G}(\bd^{\prime}), \bR_{G}(\bv^{\prime}), \bR_{G}(\bH)).
  \end{equation*}
\end{rem}

\subsection{Invariants and covariants of the elasticity tensor}
\label{subsec:Inv-Cov}

The quantities
\begin{equation*}
  \lambda=\lambda(\bE), \quad \mu=\mu(\bE), \quad \bd^{\prime}=\bd^{\prime}(\bE), \quad \bv^{\prime}= \bv^{\prime}(\bE), \quad \bH=\bH(\bE),
\end{equation*}
are \emph{covariants $\bC(\bE)$} of $\bE$~\cite{KP2000} (of degree one and respective order 0, 0, 2, 2 and 4 ).

The scalars $\lambda$ and $\mu$ are linear invariants of $\bE$, whereas $\bd^{\prime}(\bE)$, $\bv^{\prime}(\bE)$ and $\bH=\bH(\bE)$ are linear covariants of $\bE$. They satisfy the rule
\begin{equation}\label{eq:cov}
  \bC(r\star \bE) = r\star \bC(\bE), \qquad \forall r\in \SO(3),
\end{equation}
which simplifies into $I(r\star \bE)= I(\bE)$ for invariants $I(\bE)$.

There exist \emph{polynomial covariants} of higher degree, for example the quadratic covariant
\begin{equation}\label{eq:d2}
  \bd_{2}(\bH):=\bH\3dots \bH , \qquad (\emph{i.e.}, \; (\bd_{2})_{ij}=H_{ipqr}H_{pqrj}),
\end{equation}
introduced in~\cite{BKO1994}. Note that the algebra of (totally symmetric) polynomial covariants of the elasticity tensor has been defined
in~\cite{OKDD2021} and that a minimal integrity basis for the invariant algebra of $\bE$ has been derived in~\cite{OKA2017} and~\cite{OKDD2021} (it is of cardinal 294).

By definition~\eqref{eq:cov}, a covariant $\bC(\bE)$ of $\bE$ inherits the symmetry of $\bE$: the symmetry group of $\bC(\bE)$ contains  the symmetry group of $\bE$,
\begin{equation*}
  G_{\bE} \subset G_{\bC(\bE)} ,
\end{equation*}
meaning that $\bC(\bE)$ has at least the symmetry of $\bE$.

\begin{rem}\label{rem:inheritance-of-covariance}
  This property has important consequences~\cite{OKDD2021}:
  \begin{enumerate}
    \item the vector covariants $\vv(\bE)$ of a monoclinic elasticity tensor $\bE$ are all collinear,
    \item the vector covariants $\vv(\bE)$ of an elasticity tensor $\bE$ either orthotropic, tetragonal, trigonal, cubic, transversely isotropic or isotropic, all vanish:
          \begin{equation*}
            \vv(\bE)=0 \quad \forall \bE\in \strata{\DD_{2}} \cup \strata{\DD_{3}} \cup \strata{\DD_{4}} \cup \strata{\octa} \cup \strata{\OO(2)} \cup \strata{\SO(3)},
          \end{equation*}
    \item the second-order covariants $\bc(\bE)$ of an elasticity tensor either cubic or isotropic are all isotropic,
    \item the second-order covariants $\bc(\bE)$ of an elasticity tensor $\bE$ either tetragonal, trigonal or transversely isotropic, of axis $\langle \nn \rangle$, are all at least transversely isotropic of axis $\langle \nn \rangle$,
    \item the second-order covariants $\bc(\bE)$ of an orthotropic elasticity tensor $\bE$ are all at least orthotropic
          (and all of them commute with each other).
    \item the second-order covariants $\bc(\bE)$ of a triclinic elasticity tensor $\bE$ are all at least orthotropic (but the natural basis may differ from one covariant to another).
  \end{enumerate}
\end{rem}

\section{Literature on upper bounds estimates of the distance to a symmetry class}
\label{sec:Literature-Upper bounds}

Baerheim~\cite{Bae1993} has observed that the trace $\tr_{12} \bE^{a}=\Id: \bE^{a}$ of the asymmetric part~\cite{Bac1970}
\begin{equation*}
  \bE^{a}:=\bE-\bE^s,
  \qquad
  \begin{cases}
    (\bE^s)_{ijkl}=\frac{1}{3}\left(E_{ijkl}+E_{ikjl}+E_{iljk}\right),
    \\
    (\bE^{a})_{ijkl}=\frac{1}{3}\left(2 E_{ijkl}-E_{ikjl}-E_{iljk}\right)	,
  \end{cases}
\end{equation*}
of an elasticity tensor $\bE$, generically carries information related to a so-called \emph{symmetry coordinate system} of $\bE$.
This is in fact a consequence of points (4) and (5) in remark~\ref{rem:inheritance-of-covariance}.

In the case of an orthotropic tensor $\bE$, the trace $\Id:\bE^{a}$ is diagonal in the natural coordinate system of the well-known nine-dimensional orthotropic Kelvin normal form. The second-order tensor
\begin{equation}\label{eq:aStahn}
  \bt:=\Id:\bE^{a}=\frac{2}{3}\left(\bd-\bv\right),
\end{equation}
is in fact a covariant of $\bE$. As such it inherits the symmetry of $\bE$, and, generically, $\bt(\bE)$ is orthotopic if $\bE$ is orthotropic\footnote{Some degeneracy are possible for some orthotropic elasticity tensors $\bE$~\cite{OKDD2021}.}. Note that orthotropy is one of the three symmetry classes which is common to elasticity tensors and symmetric second-order tensors (see section~\ref{subsec:SymClasses}).

Starting from a given (raw, usually triclinic) elasticity tensor $\bE_{0}$ with harmonic decomposition~\eqref{eq:dec-harm-E},
\begin{equation*}
  \bE_{0}= (\lambda_{0}, \mu_{0}, \bd_{0}', \bv_{0}', \bH_{0}) ,
\end{equation*}
the fact that the second-order covariant
\begin{equation*}
  \bt_{0}=\Id:( \bE_{0}-\bE_{0}^s) =\frac{2}{3}\left(\bd_{0}-\bv_{0} \right)
\end{equation*}
carries information on the likely symmetry coordinate system, has been used by Stahn and coworkers~\cite{SMB2020} in order to define upper bounds estimates $M(\bE_{0}, \strata{G})$ of the distance of $\bE_{0}$ to a symmetry stratum $\strata{G}$. When the distance of $\bE_{0}$ to $\strata{G}$ is defined by
\begin{equation*}
  d(\bE_{0}, \strata{G})=\min_{\bE\in\strata{G}}\norm{\bE_{0}-\bE},
\end{equation*}
where $\norm{\cdot}$ is usually an $\SO(3)$-invariant norm, estimates of the distance to a symmetry class (including the 2D orthotropic estimates of~\cite[Section 5]{Via1997} and of~\cite[Corollary 3.2]{OLDK2021}, and the 3D transversely isotropic estimate of~\cite{Kli2018}) are usually obtained as
\begin{equation*}
  M(\bE_{0}, \strata{G})=\min_{\bE\in S \subset \strata{G}}\norm{\bE_{0}-\bE},
\end{equation*}
\emph{i.e.}, as the minimum over a subset $S$, astutely chosen, of the considered symmetry stratum. It satisfies thus
\begin{equation*}
  d(\bE_{0}, \strata{G}) \le M(\bE_{0}, \strata{G}).
\end{equation*}

Observe that choosing a finite subset $S$ of $N$ elasticity tensors
\begin{equation*}
  S=\set{\bE_{1}, \dotsc, \bE_{N}},
\end{equation*}
in the given symmetry stratum $\strata{G}$, lead to an estimate of $d(\bE_{0}, \strata{G})$.
The subtlety consists in choosing efficiently this set, given $\bE_{0}$.

For instance, in~\cite{SMB2020}, by the procedure recalled in \autoref{sec:Stahn}, the authors define first a finite list $\set{G_{1}, \dotsc, G_{N}}$ of representative groups in the symmetry class $[G]$, using the eigenbasis of the second-order covariant  $\bt_{0}=\bt(\bE_{0})$ (assumed to be orthotropic).  To each symmetry group $G_{n}$ is then associated an elasticity tensor $\bE_{n}$ with symmetry $G_{n}$, using the Reynolds averaging operator $\bR_{G_{n}}$ (defined for finite groups by \eqref{eq:ReyG}).

It is worth mentioning that difficulties arise when the considered elasticity symmetry class is not orthotropic, and when the second order covariant $\bt_{0}$ does not carry much information about the expected symmetry group~\cite{Bae1993,OKDD2018,ADD2020,SMB2020}. This is the case for instance for cubic symmetry, since all second-order covariants $\bc(\bE)$ are then isotropic (see \autoref{subsec:Inv-Cov}).

Another approach has been proposed by Klimeš~\cite{Kli2016,Kli2018} for transverse isotropy (symmetry class $[\OO(2)]$~\cite{FV1996}). Instead of considering that a covariant of $\bE_{0}$ carries the likely symmetry coordinate system of the optimal transversely isotropic elasticity tensor $\bE\in \strata{\OO(2)}$, this author did observe that the following relation
\begin{equation}\label{eq:KlimesT}
  \bT(\bE, \nn)=0,
  \qquad
  T_{ijkl}(\bE, \nn)=\frac{1}{4} n_{m}\left(\varepsilon_{min} E_{njkl}+\varepsilon_{mjn} E_{inkl}+\varepsilon_{mkn} E_{ijnl}+\varepsilon_{mln}
  E_{ijkn}\right),
\end{equation}
which is linear both in the transverse isotropy direction $\nn$ and $\bE$, was satisfied for each elasticity tensor $\bE$ transversely isotropic of axis $\nn$.

\begin{rem}\label{rem:n-not-cov}
  The vector $\nn$ is not a covariant of $\bE$, since all vector covariants of $\bE$ vanish when $\bE$ is transversely isotropic~\cite{OKDD2021} (see \autoref{subsec:Inv-Cov}).
\end{rem}

Klimeš did then suggest the likely transverse isotropy coordinate system of a raw tensor $\bE_{0}$, as a coordinate system with
a $z$-axis of direction a unit vector $\nn$ which minimizes
\begin{equation}\label{eq:KilimesIT}
  \min_{\norm{\nn}=1} \norm{\bT(\bE_{0}, \nn)}^{2},
\end{equation}
for the Euclidean norm. Nicely, this quadratic optimization problem has an analytical solution. Indeed, \eqref{eq:KilimesIT} can be recast as~\cite{Kli2016}
\begin{equation}\label{eq:KilimesITA}
  \min_{\norm{\nn}=1} \langle A\, \nn, \nn \rangle
\end{equation}
where $\langle \cdot, \cdot \rangle$ is the Euclidean scalar product and $A=A(\bE_{0})$ is a positive semi-definite matrix, which is moreover a quadratic second-order covariant of $\bE_{0}$ (see \autoref{sec:Klimes}). The solution of this problem is a unit eigenvector $\nn$ of $A$, associated with its smallest eigenvalue. Once the axis $\langle \nn \rangle$ of the transversely isotropic symmetry group $G$  (conjugate to $\OO(2)$) is known, an upper bound estimate
\begin{equation*}
  M(\bE_{0}, \strata{\OO(2)})= \norm{\bE_{0}-\bR_{G}(\bE_{0})}.
\end{equation*}
is obtained by standard Reynolds averaging~\cite{Kli2018} (an alternative simpler formula is provided in \autoref{sec:Klimes}).

\section{Symmetry coordinate system of a close to be cubic or orthotropic elasticity tensor}
\label{sec:symmetry-coordinate-system}

The problem now is to find a \emph{likely cubic coordinate system} (or a \emph{likely orthotropic coordinate system}) for a raw elasticity tensor $\bE_{0}$. One knows that the exact distance to cubic symmetry $d(\bE_{0}, \strata{\octa})$ is indeed solution of a quadratic optimization problem, which can be solved using the computation of a Gröbner basis~\cite{ADKP2022}. To get a fully analytical (but approximate) solution, we propose to generalize Klimeš procedure by formulating a linear equation on a tensorial variable $\vv$ which replaces the vector $\nn$ in Klimeš equation~\eqref{eq:KlimesT}. This equation is written as
\begin{equation*}
  \bT(\bC(\bE),\vv)= 0,
\end{equation*}
where $\bC(\bE)$ is a covariant of $\bE$.
\begin{itemize}
  \item In the case of transverse isotropy (Klimeš equation), $\bC(\bE) = \bE$, and $\vv=\nn$ is a vector. Solutions $\nn$ of the equation $\bT(\bE,\nn)= 0$ matches the axis of transverse isotropy of $\bE$.
  \item In the case of cubic or orthotropic symmetry, $\bC(\bE)=\bH$, where $\bH$ is the fourth-order harmonic component of $\bE$, and $\vv = \ba'$ is a symmetric second-order deviator.
\end{itemize}

Following Klimeš~\cite{Kli2016}, and considering then the measured---therefore not cubic---elasticity tensor
\begin{equation*}
  \bE_{0}=(\lambda_{0}, \mu_{0}, \bd_{0}', \bv_{0}', \bH_{0}),
\end{equation*}
the squared Euclidean norm  $\norm{\bT(\bH_{0},\ba')}^{2}$ is minimized under the constraint $\norm{\ba'}=1$ to obtain an expression of the second-order tensor $\ba'$. When $\ba'$ is orthotropic, its eigenbasis $(\ee_{i})$ defines the likely cubic coordinate system sought. We shall see that it will also define a likely orthotropic coordinate system.

\subsection{Likely cubic coordinate system of a raw elasticity tensor}

Let $\bE=(\lambda, \mu, 0, 0, \bH)$ be a cubic elasticity tensor, and let
\begin{equation}\label{eq:SxT}
  \bS_{1}\times \bS_{2} :=- (\bS_{1}\cdot \lc \cdot \bS_{2})^{s},
\end{equation}
denote the generalized cross product between two totally symmetric tensors $\bS_{1}$ and $\bS_{2}$ introduced in~\cite{OKDD2021}. Here, $\lc=(\varepsilon_{ijk})$ is the Levi-Civita tensor, a dot $\cdot$ stands for one subscript contraction, and $(\;)^{s}$ means the total symmetrization of a tensor.

It has been shown in~\cite[Appendix B]{ADD2020} that, for a given cubic harmonic fourth-order tensor $\bH\in \HH^{4}$, the linear equation
\begin{equation*}
  \bT(\bH, \ba'):=\tr ( \bH \times \ba') = 0, \qquad \ba' \in \HH^{2},
\end{equation*}
in the \emph{deviatoric second-order tensor $\ba'$} has orthotropic solutions and all of them are diagonal in the cubic coordinate system common to $\bH$ and $\bE$. See \autoref{sec:TrHxa} for the expression of the components $[\tr ( \bH \times \ba)]_{ijk}$.

\begin{rem}\label{rem:trHx1}
  We have $\bS \times \Id=0$ for every symmetric tensor $\bS$. Hence,
    \begin{equation*}
      \tr ( \bH \times \ba)=\tr ( \bH \times \ba')
    \end{equation*}
    for every symmetric second-order tensor $\ba$, where $\ba'$ denotes its deviatoric part.
\end{rem}

When the elasticity tensor $\bE_{0}=(\lambda_{0}, \mu_{0}, \bd_{0}', \bv_{0}', \bH_{0})$ is not, but close to be cubic, we expect to obtain a likely cubic basis $(\ee_{1}, \ee_{2}, \ee_{3})$ for $\bE_{0}$ where the $\ee_{i}$ are the eigenvectors of an orthotropic deviatoric second-order tensor $\ba'$, which minimizes (see \autoref{sec:eigenvalue-problem})
\begin{equation}\label{eq:minTrHxa2}
  \min_{\norm{\ba'}=1}\, \norm{\tr ( \bH_{0} \times \ba')}^{2}, \qquad \ba' \in \HH^{2}.
\end{equation}

\begin{rem}
  Note that it is very important to seek for $\ba=\ba'$ deviatoric, since, by remark \ref{rem:trHx1}, we have $\tr ( \bH_{0} \times \Id)=0$. Indeed, for each fourth-order harmonic tensor $\bH_{0}$,  the minimization problem
  \begin{equation*}
    \min_{\norm{\ba}=1}\, \norm{\tr ( \bH_{0} \times \ba)}^{2}, \qquad \ba \in \Sym^{2}
  \end{equation*}
  has for minimum $0$ and this minimum is obtained for $\ba = \pm\Id/\sqrt{3}$ (spherical), which does not furnish any information.
\end{rem}

The function to be minimized can be rewritten as
\begin{equation}\label{eq:minaA0a}
  \norm{\tr ( \bH_{0} \times \ba')}^{2}
  =\frac{9}{200} \norm{\bH_{0}}^{2} \norm{\ba'}^2 - \frac{3}{20} \ba':\bH_{0}^{2}:\ba'
  + \frac{27}{100} \bd_{2}'(\bH_{0}) :\ba^{\prime\,2},
\end{equation}
where $\bH_{0}^{2}=\bH_{0}:\bH_{0}$ and $\bd_{20}=\bd_{2}(\bH_{0})$ is the symmetric second-order covariant of $\bH_{0}$ defined by~\eqref{eq:d2}.

\begin{rem}\label{rem:B0}
  It is worth pointing out that one has furthermore $\bd_{2}'(\bH_{0}) = 0$ (\emph{i.e.}, $\bd_{2}(\bH_{0})$ is isotropic),
  when $\bH_{0}$ is cubic~\cite[Theorem 9.3]{OKDD2021}. This means that another likely cubic coordinate system, in general
  different from the one obtained by minimizing~\eqref{eq:minTrHxa2}, could also be obtained using the minimization problem
  \begin{equation*}
    \min_{\norm{\bb'}=1}\left(\frac{9}{200} \norm{\bH_{0}}^{2}\,\norm{\bb'}^{2} - \frac{3}{20} \bb' : \bH_{0}^2: \bb'\right),
    \qquad \bb' \in \HH^{2},
  \end{equation*}
  instead of~\eqref{eq:minTrHxa2}.
\end{rem}

\subsection{Likely orthotropic coordinate system of a raw elasticity tensor}
\label{subsec:Likely-Ortho-Proof}

Consider an orthotropic elasticity tensor $\bE$. In its natural basis, its Kelvin representation is written as \eqref{eq:VoigtOrtho}.
Generically, its fourth-order harmonic part $\bH$ is also orthotropic and its Kelvin representation has for expression~\cite{AKP2014}
\begin{equation}\label{eq:normal-form-D2}
  [\bH]= \left(
  \begin{array}{cccccc}
      \lambda_{2} + \lambda_{3} & -\lambda_{3}              & -\lambda_{2}              & 0               & 0               & 0               \\
      -\lambda_{3}              & \lambda_{3} + \lambda_{1} & -\lambda_{1}              & 0               & 0               & 0               \\
      -\lambda_{2}              & -\lambda_{1}              & \lambda_{2} + \lambda_{1} & 0               & 0               & 0               \\
      0                         & 0                         & 0                         & -2\,\lambda_{1} & 0               & 0               \\
      0                         & 0                         & 0                         & 0               & -2\,\lambda_{2} & 0               \\
      0                         & 0                         & 0                         & 0               & 0               & -2\,\lambda_{3}
    \end{array}
  \right) ,
\end{equation}
where $\lambda_{1},\lambda_{2},\lambda_{3}$ are three distinct real numbers.

Our goal, now, is to show that, generically, the equation
\begin{equation*}
  \tr ( \bH \times \ba')=0, \qquad \ba' \in \HH^{2},
\end{equation*}
where $\bH$ is orthotropic has orthotropic solutions and that all these solutions are diagonal in the natural basis of $\bH$. Looking for deviatoric solutions we have $a'_{33}=-a'_{11}-a'_{22}$, and, using the formulas in \autoref{sec:TrHxa}, we get
\begin{equation*}
  \begin{aligned}
    0=(\tr(\bH \times \ba'))_{111} =  & \frac{3}{10} a'_{23} (\lambda_{2}-\lambda_{3}),
    \\
    0=(\tr(\bH \times \ba'))_{112} =  & -\frac{1}{10} a'_{13} (2 \lambda_{2}+3 \lambda_{3}),
    \\
    0=(\tr(\bH \times \ba'))_{113} =  & \frac{1}{10} a'_{12} (3 \lambda_{2}+2 \lambda_{3}),
    \\
    0=(\tr(\bH \times \ba'))_{122} =  & \frac{1}{10} a'_{23} (2 \lambda_{1}+3 \lambda_{3}),
    \\
    0=(\tr(\bH \times \ba'))_{123} =  & \frac{1}{10} \left(a'_{11} (\lambda_{1}-2 \lambda_{2}+\lambda_{3})+a'_{22} (2 \lambda_{1}-\lambda_{2}-\lambda_{3})\right),
    \\
    0=(\tr(\bH \times \ba'))_{133} =  & -\frac{1}{10} a'_{23} (2 \lambda_{1}+3 \lambda_{2}),
    \\
    0=(\tr(\bH \times \ba'))_{222} =  & -\frac{3}{10}  a'_{13} (\lambda_{1}-\lambda_{3}),
    \\
    0= (\tr(\bH \times \ba'))_{223} = & -\frac{1}{10} a'_{12} (3 \lambda_{1}+2 \lambda_{3}),
    \\
    0=(\tr(\bH \times \ba'))_{233} =  & \frac{1}{10} a'_{13} (3 \lambda_{1}+2 \lambda_{2}) ,
    \\
    0=(\tr(\bH \times \ba'))_{333} =  & \frac{3}{10} a'_{12} (\lambda_{1}-\lambda_{2}) ,
  \end{aligned}
\end{equation*}
which leads to the general solution
\begin{equation*}
  \ba'=\frac{a'_{11}}{\lambda_{2}+\lambda_{3}-2 \lambda_{1}}
  \left(
  \begin{array}{ccc}
      -2 \lambda_{1}+\lambda_{2}+\lambda_{3} & 0                                     & 0                                     \\
      0                                      & \lambda_{1}-2 \lambda_{2}+\lambda_{3} & 0                                     \\
      0                                      & 0                                     & \lambda_{1}+\lambda_{2}-2 \lambda_{3} \\
    \end{array}
  \right),
\end{equation*}
which is diagonal in the natural basis of $\bH$ (and thus $\bE$). Moreover, generically $\ba'$ has three distinct eigenvalues and its three eigenvectors constitute then a natural basis for $\bH$.

This observation allows us to seek, as in the cubic case, for a likely orthotropic basis $(\ee_{i})$ for a raw (triclinic) elasticity tensor $\bE_{0}=(\lambda_{0}, \mu_{0}, \bd_{0}', \bv_{0}', \bH_{0})$, as an orthonormal eigenbasis of an orthotropic deviatoric tensor $\ba'$ which minimizes
\begin{equation}\label{eq:minTrHxa2ortho}
  \min_{\norm{\ba'}=1}\, \norm{\tr ( \bH_{0} \times \ba')}^{2}, \qquad \ba' \in \HH^{2}.
\end{equation}

\section{Reduction to an eigenvalue problem}
\label{sec:eigenvalue-problem}

In the method suggested by Klimeš, to estimate the distance to transversely isotropic symmetry class, the problem consists in minimizing the quadratic form
\begin{equation*}
  \min_{\norm{\nn}=1} \langle A_{\bE}\, \nn, \nn \rangle,
\end{equation*}
where the symmetric linear operator $A_{\bE}$ is detailed in \autoref{sec:Klimes}. It reduces therefore to the classical problem of minimizing a positive definite quadratic form $\langle A \,\vv, \vv \rangle$ defined on an Euclidean vector space over the unit sphere
\begin{equation*}
  \min_{\norm{\vv}=1}\,\langle A \,\vv, \vv \rangle,
\end{equation*}
whose solutions $\vv^{*}$ are the unit eigenvectors corresponding to the smallest eigenvalue of $A$.

In the problem we consider, which is to estimate the distance to cubic/orthotropic symmetry class, we introduce first the bilinear mapping
\begin{equation*}
  (\bH, \ba) \mapsto \tr (\bH \times \ba), \qquad \HH^{4} \times \Sym^{2} \to \HH^{3}.
\end{equation*}
To emphasize the fact that, here, $\bH$ plays the role of a parameter and $\ba$ of a variable, we recast this bilinear mapping as the linear operator
\begin{equation*}
  L_{\bH} : \ba \mapsto \tr (\bH \times \ba), \qquad \Sym^{2} \to \HH^{3}.
\end{equation*}
We have then
\begin{equation*}
  \norm{\tr (\bH \times \ba)}^{2} = \norm{L_{\bH} \,\ba}^{2} = \langle L_{\bH}\,\ba, L_{\bH}\,\ba\rangle = \langle {L_{\bH}}^{t} L_{\bH}\,\ba, \ba \rangle,
\end{equation*}
where ${L_{\bH}}^{t}$ is the transpose of the operator $L_{\bH}$ for the Euclidean metric, and
where the scalar product on $\Sym^{2}$ is given by
\begin{equation*}
  \langle \ba, \bb \rangle = \tr (\ba\bb) = \ba : \bb.
\end{equation*}
We will set thus $A_{\bH} := {L_{\bH}}^{t} L_{\bH}$, which is a positive semi-definite operator on $\Sym^{2}$. This operator can be interpreted as the fourth-order tensor (of elasticity type)
\begin{equation*}
  \bA_{\bH} = \frac{9}{200} \norm{\bH}^{2} \bJ - \frac{3}{20} \bH^{2} + \frac{27}{200} \bJ:\left( \Id \otimesbar \bd_{2}' + \bd_{2}' \otimesbar \Id \right):\bJ ,
\end{equation*}
where $\bd_{2}'$ is the deviatoric part of $\bd_{2}(\bH)=\bH\3dots \bH$ defined by~\eqref{eq:d2}, $\bH^{2}:=\bH:\bH$, and where the deviatoric projector
\begin{equation*}
  \bJ:=\bI-\frac{1}{3} \Id \otimes \Id,
\end{equation*}
satisfies $\bJ:\ba=\ba:\bJ=\ba'$ for all $\ba\in \Sym^{2}$. In particular, we have $\bA_{\bH}:\Id=\Id:\bA_{\bH}=0$.

As already stated, the minimization problem
\begin{equation*}
  \min_{\norm{\ba}=1}\, \langle A_{\bH}\,\ba, \ba \rangle, \qquad \ba \in \Sym^{2},
\end{equation*}
does not lead to any pertinent information, since the linear operator $A_{\bH}$ has always a vanishing eigenvalue (indeed, $A_{\bH}\Id=0$, for every $\bH$). Hence, in general there is no orthotropic minimum but a minimizing sequence of orthotropic tensors $(\ba_{n})$, with $\norm{\ba_{n}}=1$, converging to either $\ba= \Id/\sqrt{3}$ or $\ba= -\Id/\sqrt{3}$. This is not useful.

We need thus to remove these annoying solutions. To do so, we consider the subspace $\HH^{2}$ of $\Sym^{2}$ of deviatoric (harmonic) second-order tensors. Since $A_{\bH}\Id = 0$, $A_{\bH}$ is symmetric and $\HH^{2}$ is the orthogonal complement in $\Sym^{2}$ of the one-dimensional vector space spanned by $\Id$, the restriction $\left. A_{\bH}\right|_{\HH^{2}}$ to $\HH^{2}$  is a linear mapping from $\HH^{2}$ to $\HH^{2}$.

We shall thus consider rather the minimization problem
\begin{equation*}
  \min_{\norm{\ba'}=1}\, \langle A_{\bH}\big|_{\HH^{2}}\,\ba', \ba' \rangle, \qquad \ba' \in \HH^{2},
\end{equation*}
but with further restrictions, that we shall now explain. We note first that, \emph{in general}, the eigenspace corresponding to the vanishing eigenvalue of $A_{\bH}$ is one-dimensional and thus spanned by $\Id$. To make this statement rigorous, we restrict our study to tensors $\bH$ such that $\left. \det \left. A_{\bH}\right|_{\HH^{2}} \ne 0\right.$, which defines a generic set of tensors $\bH$ (a non-empty \emph{Zariski open set}, see \cite{Har1977,BKO1994,DADKO2019}).

If $\bH$ is generic (\textit{i.e.}, $\det \left. A_{\bH}\right|_{\HH^{2}} \ne 0$), then $\left. A_{\bH}\right|_{\HH^{2}}$ is positive definite and its eigenvalues correspond to the positive eigenvalues of $A_{\bH}$. Besides, the corresponding eigenvectors are deviatoric, since they are orthogonal to the eigenvector $\Id$ associated with the eigenvalue $\lambda=0$. Therefore, in practice, it is not necessary to calculate $\left. A_{\bH}\right|_{\HH^{2}}$. In fact, we need only to calculate the eigenvalues of $A_{\bH}$ and consider its smallest positive eigenvalue $\lambda_{min}$. An eigenvector $\ba=\ba'$ for $\lambda_{min}>0$ (which is necessary deviatoric) is thus a candidate to provide our likely cubic normal basis. To fully solve the problem, the deviatoric second-order tensor $\ba'$ has to be orthotropic, not transversely isotropic. This turns out to be generic as well, as it can be checked in the examples of sections \ref{sec:Upper bound-Cubic} and~\ref{sec:Upper bound-Ortho}.

To sum up this methodology, in practice, the minimization problem
\begin{equation*}
  \min_{\norm{\ba'}=1}\, \norm{\tr(\bH_{0}\times \ba')}^{2},
  \qquad \ba'\in \HH^{2},
\end{equation*}
for generic $\bH_{0}$ reduces to calculate the eigenvector
\begin{equation*}
  \underline \ba' =
  \begin{pmatrix}
    a'_{11} \\ a'_{22} \\ a'_{33} \\ \sqrt{2}\, a'_{23} \\  \sqrt{2}\, a'_{13} \\  \sqrt{2}\, a'_{12}
  \end{pmatrix}.
\end{equation*}
corresponding to the smallest positive eigenvalue $\lambda_{min}>0$ of the symmetric fourth-order tensor
\begin{equation*}
  \bA_{\bH_{0}} = \frac{9}{200} \norm{\bH_{0}}^{2} \bJ - \frac{3}{20} \bH_{0}^{2} + \frac{27}{200} \bJ:\left( \Id \otimesbar \bd_{20}' + \bd_{20}' \otimesbar \Id \right):\bJ,
\end{equation*}
expressed in Kelvin notation,
\begin{equation*}
  [A_{\bH_{0}}]=
  \begin{pmatrix}
    (\bA_{\bH_{0}})_{1111}             & (\bA_{\bH_{0}})_{1122}            & (\bA_{\bH_{0}})_{1133}            & \sqrt{2}\, (\bA_{\bH_{0}})_{1123} & \sqrt{2}\, (\bA_{\bH_{0}})_{1113} & \sqrt{2}\, (\bA_{\bH_{0}})_{1112} \\
    (\bA_{\bH_{0}})_{1122}             & (\bA_{\bH_{0}})_{2222}            & (\bA_{\bH_{0}})_{2233}            & \sqrt{2}\, (\bA_{\bH_{0}})_{2223} & \sqrt{2}\, (\bA_{\bH_{0}})_{2213} & \sqrt{2}\, (\bA_{\bH_{0}})_{2212} \\
    (\bA_{\bH_{0}})_{1133}             & (\bA_{\bH_{0}})_{2233}            & (\bA_{\bH_{0}})_{3333}            & \sqrt{2}\, (\bA_{\bH_{0}})_{3323} & \sqrt{2}\, (\bA_{\bH_{0}})_{3313} & \sqrt{2}\, (\bA_{\bH_{0}})_{3312} \\
    \sqrt{2}\,  (\bA_{\bH_{0}})_{1123} & \sqrt{2}\, (\bA_{\bH_{0}})_{2223} & \sqrt{2}\, (\bA_{\bH_{0}})_{3323} & 2\, (\bA_{\bH_{0}})_{2323}        & 2\, (\bA_{\bH_{0}})_{2313}        & 2\, (\bA_{\bH_{0}})_{2312}        \\
    \sqrt{2}\, (\bA_{\bH_{0}})_{1113}  & \sqrt{2}\, (\bA_{\bH_{0}})_{2213} & \sqrt{2}\, (\bA_{\bH_{0}})_{3313} & 2\, (\bA_{\bH_{0}})_{2313}        & 2\, (\bA_{\bH_{0}})_{1313}        & 2\, (\bA_{\bH_{0}})_{1312}        \\
    \sqrt{2}\,  (\bA_{\bH_{0}})_{1112} & \sqrt{2}\, (\bA_{\bH_{0}})_{2212} & \sqrt{2}\, (\bA_{\bH_{0}})_{3312} & 2\, (\bA_{\bH_{0}})_{2312}        & 2\, (\bA_{\bH_{0}})_{1312}        & 2\, (\bA_{\bH_{0}})_{1212}
  \end{pmatrix}.
\end{equation*}

\begin{rem}
  The same procedure, with $ \bA_{\bH_{0}}$ replaced by $\bB_{\bH_{0}}$ and $\ba'$ replaced by $\bb'$, applies to the minimization problem defined in remark~\eqref{rem:B0}, and recasts as
  \begin{equation}\label{eq:minB0b}
    \min_{\norm{\bb'}=1}\left(\frac{9}{200} \norm{\bH_{0}}^{2}\,\norm{\bb'}^{2} - \frac{3}{20} \bb' : \bH_{0}^2: \bb'\right)
    =\min_{\norm{\bb'}=1}\, \bb' :  \bB_{\bH_{0}}\big|_{\HH^{2}} : \bb', \qquad \bb' \in \HH^{2},
  \end{equation}
  where we seek for $\bb'$ deviatoric and orthotropic, and where $\bB_{\bH_{0}}\big|_{\HH^{2}}$ is the restriction to $\HH^{2}$ of the symmetric operator
  \begin{equation*}
    \bB_{\bH_{0}}= \frac{9}{200} \norm{\bH_{0}}^{2}\, \bJ - \frac{3}{20} \bH_{0}^2,
  \end{equation*}
  defined on $\Sym^{2}$ (and which satisfies $\bB_{\bH_{0}}:\Id=\Id:\bB_{\bH_{0}}=0$).
\end{rem}

\section{Upper bounds estimates of the distance to cubic elasticity}
\label{sec:Upper bound-Cubic}

Let $\bE_{0}$ be a given (measured) elasticity tensor, triclinic, with harmonic decomposition
\begin{equation*}
  \bE_{0}=(\lambda_{0}, \mu_{0}, \bd_{0}', \bv_{0}', \bH_{0}).
\end{equation*}
Some upper bounds estimates of the distance $d(\bE_{0}, [\octa])$ of $\bE_{0}$ to elasticity cubic symmetry, denoted here by $M(\bE_{0}, [\octa])$, have been obtained by Stahn and coworker in~\cite{SMB2020} (see \autoref{sec:Stahn}) and recently in~\cite[Appendix B]{ADKP2022}, denoted here by $\Delta_{\bd_{2}}(\bE_{0}, [\octa])$) (see \eqref{eq:DeltaNousd2}). Both estimates use a symmetric second-order covariant of $\bE_0$. In~\cite{SMB2020}, it is
\begin{equation*}
  \bt_{0}:=\frac{2}{3}( \bd_{0}-\bv_{0}),
\end{equation*}
and in \cite{ADKP2022}, it is
\begin{equation*}
  \bd_{20}:=\bd_2(\bH_{0})=\bH_{0}\3dots \bH_{0}.
\end{equation*}
The estimates $M(\bE_{0}, [\octa])/\norm{\bE_{0}}$ and $\Delta_{\bd_{2}}(\bE_{0}, [\octa])/\norm{\bE_{0}}$ of the relative distance to cubic symmetry are approximately 0.34-0.33 when evaluated for the material tested by François et al~\cite{FGB1998} (see~\cite{ADKP2022} and \autoref{tab:comp-upper}). These upper bounds estimates are not very accurate since the relative distance $d(\bE_{0}, [\octa])/\norm{\bE_{0}}$ has been computed to be 0.104 in that case~\cite{FGB1998}. The explanation is that the material considered is a single crystal superalloy (of CSMX-4 type) with a cubic  Ni-based microstructure. Its mechanical behavior is expected to be cubic (experimentally, close to be cubic). As recalled in section \ref{subsec:Inv-Cov}, all the second-order covariants of its elasticity tensor, including $\bt_{0}$ and $\bd_{20}$, are expected to be isotropic (experimentally close to be isotropic) and therefore not to any carry information about the cubic coordinate system.

The cubic estimate $\Delta_{\bd_{2}}(\bE_{0}, [\octa])$ is defined in~\cite{ADKP2022} as
\begin{equation}\label{eq:DeltaNousd2}
  \Delta_{\bd_{2}}(\bE_{0}, [\octa]) = \norm{\bE_{0}- \bE},
  \qquad
  \bE=2\mu_{0} \bI +\lambda_{0} \Id \otimes \Id+ \left(\bC_{\bd_{20}}::\bH_0\right) \bC_{\bd_{20}}\in \strata{\octa},
\end{equation}
thanks to the introduction of a cubic harmonic fourth-order tensor
\begin{equation*}
  \bC_{\ba}:=
  \sqrt{\frac{15}{2}}\; \frac{\left(\left( \ba^{\, 2} \times \ba\right)\cdot \left( \ba^{\, 2} \times \ba\right)\right)'}{\norm{\ba^{\, 2} \times \ba}^2} \in \HH^{4} ,
\end{equation*}
with $\norm{\bC_{\ba}}=1$ and where $(\cdot)'$ means the leading harmonic part,
\begin{equation*}
  \left(\left( \ba^{\, 2} \times \ba\right)\cdot \left( \ba^{\, 2} \times \ba\right)\right)'=
  \left( \ba^{\, 2} \times \ba\right)\cdot \left( \ba^{\, 2} \times \ba\right)-\frac{1}{15}\norm{\ba^{\, 2} \times \ba}^2\left( 3 \,\bI - \Id \otimes \Id \right)
  \in \HH^{4}.
\end{equation*}
It is built using the assumed orthotropic second-order tensor $\ba=\bd_{20}$ and where the component of the generalized cross product $\ba \times \bb$ of two symmetric second-order tensors defined by~\eqref{eq:SxT}, are given in \autoref{sec:TrHxa}. Note that
\begin{equation}\label{eq:Bif-a-isoT}
  \norm{\ba^{\, 2} \times \ba}^2=\frac{1}{12}\left( \left(\tr(\ba^{\prime\, 2})\right)^3-6 \left(\tr(\ba^{\prime\,3})\right)^2\right).
\end{equation}
and that $\ba^{\, 2} \times \ba \neq 0$ when the symmetric second-order tensor $\ba$ is orthotropic. Indeed, if $\ba=\text{diag}[a_{1}, a_{2}, a_{3}]$ is diagonal in some basis $(\ee_{i})$, one has~\cite{OKDD2021}
\begin{equation}\label{eq:a2xa}
  \ba^{\, 2} \times \ba=(a_{1}-a_{2}) (a_{1}-a_{3}) (a_{2}-a_{3}) \, \ee_{1} \odot \ee_{2}\odot \ee_{3},
\end{equation}
where $\odot$ is the symmetrized tensor product.

We get thus the following Kelvin matrix representations, in the same basis  $(\ee_{i})$,
\begin{equation}\label{eq:RelationPa}
  \left[\frac{\left( \ba^{\, 2} \times \ba\right)\cdot \left( \ba^{\, 2} \times \ba\right)}{\norm{\ba^{\, 2} \times \ba}^2}\right]
  = \frac{1}{3}\left(
  \begin{array}{cccccc}
      0 & 0 & 0 & 0 & 0 & 0 \\
      0 & 0 & 0 & 0 & 0 & 0 \\
      0 & 0 & 0 & 0 & 0 & 0 \\
      0 & 0 & 0 & 1 & 0 & 0 \\
      0 & 0 & 0 & 0 & 1 & 0 \\
      0 & 0 & 0 & 0 & 0 & 1 \\
    \end{array}
  \right),
\end{equation}
and
\begin{equation}\label{eq:Ca}
  [\bC_{\ba}]=\frac{1}{\sqrt{30}}
  \left(
  \begin{array}{cccccc}
      -2 & 1  & 1  & 0 & 0 & 0 \\
      1  & -2 & 1  & 0 & 0 & 0 \\
      1  & 1  & -2 & 0 & 0 & 0 \\
      0  & 0  & 0  & 2 & 0 & 0 \\
      0  & 0  & 0  & 0 & 2 & 0 \\
      0  & 0  & 0  & 0 & 0 & 2 \\
    \end{array}
  \right).
\end{equation}

\begin{rem}[Geometry of $\bC_{\ba}$]
  The symmetry group of the cubic tensor $\bC_{\ba}$ is denoted $G_{\bC_{\ba}}$ (\emph{i.e.}, $g\star \bC_{\ba} = \bC_{\ba} \; \forall g \in G_{\bC_{\ba}}$, see \autoref{sec:InvCovEla}). Since, furthermore, $\bC_{\ba}$ is a covariant of $\ba^{2}\times \ba$, which is itself a covariant of $\ba$, we get thus
  \begin{equation*}
    G_{\ba}\subset G_{\ba^{2}\times \ba}\subset  G_{\bC_{\ba}}.
  \end{equation*}
  In a natural basis $(\ee_{i})$ for $\ba$, in which $\ba$ is diagonal, we have $G_{\ba}=\DD_{2}$, and $\ba^{\, 2} \times \ba$ is of the form \eqref{eq:a2xa}. Moreover, $G_{\ba^{2}\times \ba}=\tetra$, where $\tetra$ is the tetrahedral group and $G_{\bC_{\ba}}=\octa$, where $\octa$ is the octahedral (cubic) group. When expressed in the basis $(\ee_{i})$, $\bC_{\ba}$ is fixed by $\octa$ (\textit{i.e.}, $\bC_{\ba}\in \Fix(\octa)$) is thus in its cubic normal form for an harmonic fourth-order tensor~\cite{AKP2014}. In another basis, $G_{\bC_{\ba}}= r \octa\, r^{-1}$, for some rotation $r$, and $\bC_{\ba}\in \Fix(G_{\bC_{\ba}})= \Fix(r \octa\, r^{-1})$ is fixed by a conjugate of $\octa$.
\end{rem}

\begin{rem}[Reynolds averaging / Orthogonal projection on $\Fix(G_{\bC_{\ba}})$]\label{rem:ReynolodsCubic}
  In \eqref{eq:DeltaNousd2}, the cubic harmonic tensor $\bH=\left(\bC_{\ba}::\bH_0\right) \bC_{\ba}$ and the elasticity tensor $\bE$
  are the orthogonal projections of the harmonic tensor $\bH_{0}$ and $\bE_{0}$ on the respective fixed point sets $\Fix(G_{\bC_{\ba}})$, of the harmonic fourth-order tensors and the elasticity tensors. By uniqueness of the orthogonal projection, $\bH$ and $\bE$ correspond also to Reynolds averaging (defined in \autoref{subsec:SymClasses})
  \begin{equation*}
    \bH=\bR_{G_{\bC_{\ba}}}(\bH_{0})= \left(\bC_{\ba}::\bH_0\right) \bC_{\ba} ,
    \qquad
    \bE=\bR_{G_{\bC_{\ba}}}(\bE_{0})=(\lambda_{0}, \mu_{0},0, 0, \bR_{G_{\bC_{\ba}}}(\bH_{0}) ).
  \end{equation*}
\end{rem}

Since the harmonic fourth-order tensor $\bC_{\ba}$ is cubic, it is independent from the \emph{distinct} values of the $a_{i}$.
One can choose any orthotropic second-order tensor $\ba$ (for instance other than $\bd_{20}$) to get a cubic tensor $\bE$ and to define an (other) upper bound
estimate of $d(\bE_{0}, [\octa])$, as
\begin{equation}\label{eq:DeltaNous1}
  \Delta_{\ba}(\bE_{0}, [\octa]) = \norm{\bE_{0}- \bE},
  \qquad
  \bE=2\mu_{0} \bI +\lambda_{0} \Id \otimes \Id+ \left(\bC_{\ba}::\bH_0\right) \bC_{\ba}\in \Sigma{[\octa]}.
\end{equation}
We have furthermore, by remark~\ref{rem:normE2}, the following invariant formula for the Euclidean norm,
\begin{equation}\label{eq:DeltaNous2}
  \Delta_{\ba}(\bE_{0}, [\octa])= \sqrt{\frac{2}{21} \norm{\bd_{0}^{\prime}+2 \bv_{0}^{\prime}}^{2} +\frac{4}{3}\norm{\bd_{0}^{\prime}- \bv_{0}^{\prime}}^{2}
    + \norm{\bH_0}^2 - \left(\bC_{\ba}::\bH_0\right)^2\;}\;.
\end{equation}
The \emph{cubic upper bound estimate} in~\cite{SMB2020} is then simply recovered as
\begin{equation*}
  M(\bE_{0}, [\octa])=\Delta_{\bt_{0}}(\bE_{0}, [\octa]),
\end{equation*}
by setting
\begin{equation*}
  \ba=\bt_{0}=\frac{2}{3}\left(\bd_{0}-\bv_{0}\right).
\end{equation*}

\begin{rem}
  Expression~\eqref{eq:DeltaNous1} is valid for any norm (not necessarily $\SO(3)$-invariant see~\cite{MN2006,MGD2019}), whereas expression~\eqref{eq:DeltaNous2} only applies to the Euclidean norm.
\end{rem}

The normal form $\bE_{\octa}$ (defined by~\eqref{eq:VoigtCubic}) of the cubic estimate $\bE$ is obtained directly from~\eqref{eq:Ca} with
\begin{align*}
  (\bE_{\octa})_{1111} & = 2\mu_{0} +\lambda_{0}-\frac{2}{\sqrt{30}} \bC_{\ba}::\bH_0,
  \\
  (\bE_{\octa})_{1122} & = \lambda_{0}+\frac{1}{\sqrt{30}} \bC_{\ba}::\bH_0,
  \\
  (\bE_{\octa})_{1212} & =\mu_{0} +\frac{1}{\sqrt{30}} \bC_{\ba}::\bH_0.
\end{align*}

In the following examples, we systematically compare the upper bounds estimates $\Delta_{\ba}(\bE_{0}, [\octa])$  (resp. the relative estimates $\Delta_{\ba}(\bE_{0}, [\octa])/\norm{\bE_0}$) to the distance $d(\bE_{0}, [\octa])$ (resp. the relative distance
$d(\bE_{0}, [\octa])/\norm{\bE_0}$) of a raw elasticity tensor $\bE_{0}$ to cubic symmetry, for the four different choices:

\begin{itemize}
  \item $\ba=\bt_{0}$ (Stahn and coworkers estimate of the distance to cubic symmetry~\cite{SMB2020}),
  \item $\ba=\bd_{20}$ (estimate of the distance to cubic symmetry of ref.~\cite {ADKP2022}),
  \item $\ba = \ba'$, obtained by the minimization of $\norm{\tr(\bH_{0}\times \ba')}^{2}=\ba': \bA_{\bH_{0}}\big|_{\HH^{2}}:\ba'$ (see \autoref{sec:eigenvalue-problem}).
  \item $\ba=\bb'$, obtained by the minimization of $\bb': \bB_{\bH_{0}}\big|_{\HH^{2}}:\bb'$ (see \autoref{sec:eigenvalue-problem}).
\end{itemize}
The best upper bound estimate of $d(\bE_{0}, [\octa])$ will, then, be the minimum minimorum
\begin{equation*}
  \Delta_{\text{opt}}(\bE_{0}, [\octa])=\min\left(\Delta_{\bt_{0}}(\bE_{0}, [\octa]), \Delta_{\bd_{2}}(\bE_{0}, [\octa]), \Delta_{\ba'}(\bE_{0}, [\octa]),\Delta_{\bb'}(\bE_{0}, [\octa])\right).
\end{equation*}

\begin{rem}\label{rem:many-cov2}
  Any other second-order covariant $\bc(\bE_{0})$ of the elasticity tensor $\bE_{0}$ can be added to the list
  $\set{\bt_{0}, \bd_{20}, \ba', \bb'}$, such as~\cite{OKDD2021}
  \begin{align*}
    \bd_{0},                   &  & \bv_{0},         &  & \bd_{0}^{2},         &  & \bv_{0}^{2},          &  & (\bd_{0} \bv_{0})^{s},         \\
    \bH_{0}:\bd_{0},           &  & \bH_{0}:\bv_{0}, &  & \bH_{0}:\bd_{0}^{2}, &  & \bH_{0}:\bv_{0}^{2} , &  & \bH_{0}:(\bd_{0} \bv_{0})^{s}, \\
    \bc_{3}=\bH_{0}: \bd_{20}, &  &
    \bc_{4}=\bH_{0}: \bc_{3},  &  &
    \bc_{5}=\bH_{0}: \bc_{4},  &  & \dotsc           &  &
  \end{align*}
\end{rem}

\subsection{Example of Ni-based superalloy}
\label{subsec:E0MF-cubic}

Consider the elasticity tensor (in Kelvin representation)
\begin{equation}\label{eq:E0}
  [\bE_{0}]=
  \begin{pmatrix}
    243           & 136           & 135           & 22\,\sqrt{2}  & 52\,\sqrt{2}  & -17\,\sqrt{2} \\
    136           & 239           & 137           & -28\,\sqrt{2} & 11\,\sqrt{2}  & 16\,\sqrt{2}  \\
    135           & 137           & 233           & 29\,\sqrt{2}  & -49\,\sqrt{2} & 3\,\sqrt{2}   \\
    22\,\sqrt{2}  & -28\,\sqrt{2} & 29\,\sqrt{2}  & 133 \cdot 2   & -10 \cdot 2   & -4 \cdot 2    \\
    52\,\sqrt{2}  & 11\,\sqrt{2}  & -49\,\sqrt{2} & -10 \cdot 2   & 119 \cdot 2   & -2 \cdot 2    \\
    -17\,\sqrt{2} & 16\,\sqrt{2}  & 3\,\sqrt{2}   & -4 \cdot 2    & -2 \cdot 2    & 130 \cdot 2
  \end{pmatrix} \;\text{ GPa},
\end{equation}
measured by François and coworkers~\cite{FGB1998} for a single crystal Ni-based superalloy with a so-called cubic $\gamma/\gamma'$ microstructure~\cite{FS1987,PT2006,Ree2006}. Its harmonic components $\lambda_{0}, \mu_{0}, \bd_{0}', \bv_{0}', \bH_{0}$ are given in Appendix \ref{subsec:FGB1998}. We get
\begin{equation*}
  \bt_{0}=\frac{2}{3}(\bd_{0}-\bv_{0})=\left(
  \begin{array}{ccc}
      14.67 & 8.67 & 10    \\
      8.67  & 6.67 & 16    \\
      10    & 16   & 13.33 \\
    \end{array}
  \right)
  ,\quad
  \bd_{20}=\bH_{0}\3dots \bH_{0} =10^{3}\, \left(
  \begin{array}{ccc}
      19.43 & -1.44 & -1.06 \\
      -1.44 & 15.84 & 0.83  \\
      -1.06 & 0.83  & 22.62 \\
    \end{array}
  \right),
\end{equation*}
respectively in GPa and in GPa$^{2}$, and
\begin{equation*}
  [\bC_{\bt_{0}}]=10^{-3}\,\left(
  \begin{array}{cccccc}
      18             & 38            & -56           & -185\,\sqrt{2} & 140\,\sqrt{2} & 94\,\sqrt{2}  \\
      38             & 122           & -161          & 169\,\sqrt{2}  & -66\,\sqrt{2} & -38\,\sqrt{2} \\
      -56            & -161          & 217           & 15\,\sqrt{2}   & -75\,\sqrt{2} & -56\,\sqrt{2} \\
      -185\,\sqrt{2} & 169\,\sqrt{2} & 15\,\sqrt{2}  & -161 \cdot 2   & -56 \cdot 2   & -66 \cdot 2   \\
      140\,\sqrt{2}  & -66\,\sqrt{2} & -75\,\sqrt{2} & -56 \cdot 2    & -56 \cdot 2   & -185 \cdot 2  \\
      94\,\sqrt{2}   & -38\,\sqrt{2} & -56\,\sqrt{2} & -66 \cdot 2    & -185 \cdot 2  & 38 \cdot 2    \\
    \end{array}
  \right)
  ,
\end{equation*}
\begin{equation*}
  [\bC_{\bd_{20}}]= 10^{-3}\, \left(
  \begin{array}{cccccc}
      -45            & 45             & 0              & 62\,\sqrt{2}   & -203\,\sqrt{2} & 154\,\sqrt{2}  \\
      45             & -192           & 148            & 50\,\sqrt{2}   & 2\,\sqrt{2}    & -226\,\sqrt{2} \\
      0              & 148            & -147           & -112\,\sqrt{2} & 201\,\sqrt{2}  & 73\,\sqrt{2}   \\
      62\,\sqrt{2}   & 50\,\sqrt{2}   & -112\,\sqrt{2} & 148 \cdot 2    & 73 \cdot 2     & 2 \cdot 2      \\
      -203\,\sqrt{2} & 2\,\sqrt{2}    & 201\,\sqrt{2}  & 73 \cdot 2     & 0              & 62 \cdot 2     \\
      154\,\sqrt{2}  & -226\,\sqrt{2} & 73\,\sqrt{2}   & 2 \cdot 2      & 62 \cdot 2     & 45 \cdot 2     \\
    \end{array}
  \right) .
\end{equation*}
so that
\begin{equation*}
  \bC_{\bt_{0}}::\bH_{0}=-49.32
  \;\text{ GPa},
  \qquad
  \bC_{\bd_{20}}::\bH_{0}=  -62.64 \;\text{ GPa}.
\end{equation*}
The second-order tensors $\ba'$, minimizing~\eqref{eq:minTrHxa2} and $\bb'$, minimizing~\eqref{eq:minB0b}, have for Kelvin representation the eigenvectors $\underline \ba'$ and $\underline \bb'$, corresponding to the smallest positive eigenvalue of the $6\times 6$ matrices $[\bA_{\bH_{0}}]$ and $[\bB_{\bH_{0}}]$ introduced in \autoref{sec:symmetry-coordinate-system} (and given in Appendix \ref{subsec:FGB1998}). We have
\begin{equation}\label{eq:abFGB1998}
  \ba'=\left(
  \begin{array}{ccc}
      -0.0556 & 0.0505 & -0.2156 \\
      0.0505  & 0.6837 & 0.1360  \\
      -0.2156 & 0.1360 & -0.6281 \\
    \end{array}
  \right)
  ,\qquad
  \bb'=\left(
  \begin{array}{ccc}
      0.5310  & 0.0217 & -0.3192 \\
      0.0217  & 0.1582 & 0.0816  \\
      -0.3192 & 0.0816 & -0.6892 \\
    \end{array}
  \right)
\end{equation}
and thus (in Kelvin notation)
\begin{equation*}
  [\bC_{\ba'}]=10^{-3}\,
  \left(
  \begin{array}{cccccc}
      -198          & 180           & 18             & 7\,\sqrt{2}   & 220\,\sqrt{2}  & 8\,\sqrt{2}   \\
      180           & -346          & 166            & -85\,\sqrt{2} & -6\,\sqrt{2}   & -35\,\sqrt{2} \\
      18            & 166           & -184           & 78\,\sqrt{2}  & -214\,\sqrt{2} & 27\,\sqrt{2}  \\
      7\,\sqrt{2}   & -85\,\sqrt{2} & 78\,\sqrt{2}   & 166\cdot 2    & 27\cdot 2      & -6\cdot 2     \\
      220\,\sqrt{2} & -6\,\sqrt{2}  & -214\,\sqrt{2} & 27\cdot 2     & 18\cdot 2      & 7\cdot 2      \\
      8\,\sqrt{2}   & -35\,\sqrt{2} & 27\,\sqrt{2}   & -6\cdot 2     & 7\cdot 2       & 180\cdot 2    \\
    \end{array}
  \right)
  ,
\end{equation*}
\begin{equation*}
  [\bC_{\bb'}]=10^{-3}\,
  \left(
  \begin{array}{cccccc}
      -267          & 182           & 85             & 5\,\sqrt{2}   & 188\,\sqrt{2}  & -2\,\sqrt{2}  \\
      182           & -350          & 169            & -79\,\sqrt{2} & -3\,\sqrt{2}   & -16\,\sqrt{2} \\
      85            & 169           & -254           & 74\,\sqrt{2}  & -184\,\sqrt{2} & 18\,\sqrt{2}  \\
      5\,\sqrt{2}   & -79\,\sqrt{2} & 74\,\sqrt{2}   & 169\cdot 2    & 18\cdot 2      & -3\cdot 2     \\
      188\,\sqrt{2} & -3\,\sqrt{2}  & -184\,\sqrt{2} & 18\cdot 2     & 85\cdot 2      & 5\cdot 2      \\
      -2\,\sqrt{2}  & -16\,\sqrt{2} & 18\,\sqrt{2}   & -3\cdot 2     & 5\cdot 2       & 182\cdot 2    \\
    \end{array}
  \right) ,
\end{equation*}
so that
\begin{equation*}
  \bC_{\ba'}::\bH_{0}= 218.29
  \;\text{ GPa},
  \qquad
  \bC_{\bb'}::\bH_{0}=
  226.47
  \;\text{ GPa}.
\end{equation*}

The upper bound estimates~\eqref{eq:DeltaNous2} as well as the corresponding relative estimates
\begin{equation*}
  \frac{M(\bE_{0}, [\octa])}{\norm{\bE_{0}}}=\frac{\Delta_{\bt_{0}}(\bE_{0}, [\octa])}{\norm{\bE_{0}}}, \quad
  \frac{\Delta_{\bd_{20}}(\bE_{0}, [\octa])}{\norm{\bE_{0}}}, \quad
  \frac{\Delta_{\ba'}(\bE_{0}, [\octa])}{\norm{\bE_{0}}}, \quad
  \frac{\Delta_{\bb'}(\bE_{0}, [\octa])}{\norm{\bE_{0}}},
\end{equation*}
are compared in \autoref{tab:comp-upper}, respectively to the exact distance $d(\bE_{0}, [\octa])=74.13$ GPa and to the relative distance
$d(\bE_{0}, [\octa])/\norm{\bE_0}=0.1039$ (computed in~\cite{FGB1998,ADKP2022}). The minimum minimorum is also given. The determination of a likely cubic coordinate system by the minimization problems~\eqref{eq:minTrHxa2} and~\eqref{eq:minB0b}, which avoid the use of second-order covariants of $\bE_{0}$, is found to be accurate. The best upper bound estimate of $d(\bE_{0}, [\octa])$ is here $\Delta_{\bb'}(\bE_{0}, [\octa])=97.8$ GPa, so that $\Delta_{\bb'}(\bE_{0}, [\octa])/\norm{\bE_{0}}=0.1371$, with $\bb'$ the deviatoric solution of $\min_{\norm{\bb'}=1}\, \bb' : \bB_{\bH_{0}}\big|_{\HH^{2}} : \bb'$ where
\begin{equation*}
  \bB_{\bH_{0}}= \frac{9}{200} \norm{\bH_{0}}^{2}\, \bJ - \frac{3}{20} \bH_{0}^2 .
\end{equation*}

\begin{table}[h]
  \begin{center}
    \setlength{\arraycolsep}{1pt}
    \begin{tabular}{c|ccccc|c}
      \toprule
      $d(\bE_{0}, [\octa])$ &                    & $M=\Delta_{\bt_{0}}$ & $ \Delta_{\bd_{20}}$ & $ \Delta_{\ba'}$ & $ \Delta_{\bb'}$ & $ \Delta_{\text{opt}}= \Delta_{\bb'}$ \\
      \midrule
      74.13                 & Estimate (GPa):    & 241.7                & 238.6                & 114.9            & 97.8             & 97.8                                  \\
      \midrule
      0.1039                & Relative estimate: & 0.3388               & 0.3344               & 0.1610           & 0.1371           & 0.1371                                \\
      \bottomrule
    \end{tabular}
    \caption{Comparison of upper bounds estimates of the distance to cubic elasticity $d(\bE_{0}, [\octa])$ for Ni-based single crystal superalloy.}
    \label{tab:comp-upper}
  \end{center}
\end{table}
Finally, the cubic elasticity tensor corresponding to $\Delta_{\bb'}(\bE_{0}, [\octa])=97.8$ GPa
is (in Kelvin representation and in the basis in which is expressed $[\bE_{0}]$)
\begin{equation*}
  [\bE]=
  \left(
  \begin{array}{cccccc}
      238.76          & 146.72           & 124.86           & 1.18\,\sqrt{2}   & 42.46\,\sqrt{2}  & -0.41\,\sqrt{2} \\
      146.72          & 219.91           & 143.71           & -17.89\,\sqrt{2} & -0.71\,\sqrt{2}  & -3.66\,\sqrt{2} \\
      124.86          & 143.71           & 241.76           & 16.72\,\sqrt{2}  & -41.75\,\sqrt{2} & 4.07\,\sqrt{2}  \\
      1.18\,\sqrt{2}  & -17.89\,\sqrt{2} & 16.72\,\sqrt{2}  & 135.04\cdot 2    & 4.07\cdot 2      & -0.71\cdot 2    \\
      42.46\,\sqrt{2} & -0.71\,\sqrt{2}  & -41.75\,\sqrt{2} & 4.07\cdot 2      & 116.19\cdot 2    & 1.18\cdot 2     \\
      -0.41\,\sqrt{2} & -3.66\,\sqrt{2}  & 4.07\,\sqrt{2}   & -0.71\cdot 2     & 1.18\cdot 2      & 138.05\cdot 2   \\
    \end{array}
  \right)
  \;\text{ GPa}.
\end{equation*}

\subsection{Academic example of a close to be orthotropic material}
\label{subsec:E0SMB-cubic}

Consider now the elasticity tensor studied in~\cite{SMB2020} (directly given in Kelvin notation, the units having been suppressed for brevity),
\begin{equation}\label{eq:E0Stahn}
  [\bE_{0}]=
  \left(
  \begin{array}{cccccc}
      9.35  & 5.81  & -0.20 & 5.1   & 4.06  & -2.51 \\
      5.81  & 11.09 & 2.67  & -0.83 & -3.92 & 2.79  \\
      -0.20 & 2.67  & 11.01 & -0.1  & -2.95 & 0.57  \\
      5.1   & -0.83 & -0.1  & 8.13  & -1.16 & 0.8   \\
      4.06  & -3.92 & -2.95 & -1.16 & 7.94  & 2.01  \\
      -2.51 & 2.79  & 0.57  & 0.8   & 2.01  & 8.13  \\
    \end{array}
  \right).
\end{equation}

The results are given in \autoref{tab:comp-upper-Stahn}. The estimates obtained are  this time very close to each other and to the exact distance to cubic elasticity $d(\bE_{0}, [\octa])=11.4551$ GPa (computed using a polynomial optimization method~\cite{ADKP2022}). This is not so surprising since $\bE_{0}$ is close to be orthotropic~\cite{SMB2020} (see section \ref{subsec:SMB-ortho}), so that its second-order covariants carry then accurate information on the material symmetry coordinate system.

We only provide the details concerning the upper bound estimate $\Delta_{\ba'}(\bE_{0}, [\octa])=11.4551$ GPa, which is optimal, and obtained by the minimization of
$\norm{\tr ( \bH_{0} \times \ba')}^{2}=\left. \ba' : \bA_{0}\big|_{\HH^{2}} : \ba'\right.$, where (see Appendix \ref{subsec:SMB2020})
\begin{equation}\label{eq:A0}
  \bA_{0}= \frac{9}{200} \norm{\bH_{0}}^{2} \bJ - \frac{3}{20} \bH_{0}^{2}
  +\frac{27}{200} \bJ:\left( \Id \otimesbar \bd_{20}'+ \bd_{20}' \otimesbar \Id \right):\bJ.
\end{equation}
We get
\begin{equation*}
  \ba'=\left(
  \begin{array}{ccc}
    0.5009  & 0.2749  & -0.4724 \\
    0.2749  & -0.2896 & -0.1076 \\
    -0.4724 & -0.1076 & -0.2113 \\
  \end{array}
  \right)
  ,
\end{equation*}
and (in Kelvin notation)
\begin{equation*}
  [\bC_{\ba'}]=10^{-3}\,
  \left(
  \begin{array}{cccccc}
      -21            & 113            & -93            & 122\,\sqrt{2}  & 166\,\sqrt{2}  & -123\,\sqrt{2} \\
      113            & -134           & 21             & -242\,\sqrt{2} & 22\,\sqrt{2}   & 66\,\sqrt{2}   \\
      -93            & 21             & 72             & 12\,\sqrt{2} 0 & -188\,\sqrt{2} & 58\,\sqrt{2}   \\
      122\,\sqrt{2}  & -242\,\sqrt{2} & 120\,\sqrt{2}  & 21\cdot 2      & 58\cdot 2      & 22\cdot 2      \\
      166v           & 22\,\sqrt{2}   & -188\,\sqrt{2} & 58\cdot 2      & -93\cdot 2     & 122\cdot 2     \\
      -123\,\sqrt{2} & 66\,\sqrt{2}   & 58\,\sqrt{2}   & 22\cdot 2      & 122\cdot 2     & 113\cdot 2     \\
    \end{array}
  \right),
\end{equation*}
so that
\begin{equation*}
  \bC_{\ba'}::\bH_{0}= 9.078
  \;\text{ GPa}.
\end{equation*}

\begin{table}[h]
  \begin{center}
    \setlength{\arraycolsep}{1pt}
    \begin{tabular}{c|ccccc|c}
      \toprule
      $d(\bE_{0}, [\octa])$ &          & $M=\Delta_{\bt_{0}}$ & $ \Delta_{\bd_{20}}$ & $ \Delta_{\ba'}$ & $ \Delta_{\bb'}$ & $\Delta_{\text{opt}}= \Delta_{\ba'}$ \\
      \midrule
      11.4551               &
      Estimate:             & 11.4573  & 11.4552              & 11.4551              & 14.3961          & 11.4551                                                 \\
      \midrule
      0.409081              &
      Relative estimate:    & 0.409162 & 0.409086             & 0.409083             & 0.514110         & 0.409083                                                \\
      \bottomrule
    \end{tabular}
    \caption{Comparison of upper bounds estimates of the distance to cubic elasticity $d(\bE_{0}, [\octa])$ for an academic elasticity tensor.}
    \label{tab:comp-upper-Stahn}
  \end{center}
\end{table}

The cubic elasticity tensor corresponding to $\Delta_{\ba'}(\bE_{0}, [\octa])$
is (in Kelvin representation, and in the basis in which is expressed $[\bE_{0}]$)
\begin{equation*}
  [\bE]=\left(
  \begin{array}{cccccc}
      10.07 & 3.    & 2.94  & 0.19  & -0.29 & -0.14 \\
      3.    & 10.49 & 2.52  & 0.06  & -0.23 & 0.77  \\
      2.94  & 2.52  & 10.55 & -0.25 & 0.52  & -0.63 \\
      0.19  & 0.06  & -0.25 & 7.58  & -0.88 & -0.33 \\
      -0.29 & -0.23 & 0.52  & -0.88 & 8.42  & 0.27  \\
      -0.14 & 0.77  & -0.63 & -0.33 & 0.27  & 8.54  \\
    \end{array}
  \right)
  \;\text{ GPa}.
\end{equation*}

\subsection{Example of Vosges sandstone}
\label{subsec:E0Francois-cubic}

We end this set of examples by considering the elasticity tensor identified by François~\cite[Chap. 4]{Fra1995} using the ultrasonic measurement of Arts~\cite{Art1993} on a Vosges sandstone specimen,
\begin{equation}\label{eq:E0Francois}
  [\bE_{0}]=
  \left(
  \begin{array}{cccccc}
      12.2           & -2.2           & 1.9           & 0.9\,\sqrt{2} & 0.8\,\sqrt{2}  & -0.5\,\sqrt{2} \\
      -2.2           & 13.            & 2.9           & 0.2\,\sqrt{2} & -0.3\,\sqrt{2} & 0.4\,\sqrt{2}  \\
      1.9            & 2.9            & 13.9          & 0             & 0              & 0.1\,\sqrt{2}  \\
      0.9\,\sqrt{2}  & 0.2\,\sqrt{2}  & 0             & 4.\cdot 2     & 1.2\cdot 2     & 0              \\
      0.8\,\sqrt{2}  & -0.3\,\sqrt{2} & 0             & 1.2\cdot 2    & 5.4\cdot 2     & 0              \\
      -0.5\,\sqrt{2} & 0.4\,\sqrt{2}  & 0.1\,\sqrt{2} & 0             & 0              & 5.4\cdot 2     \\
    \end{array}
  \right)
  \;\text{ GPa}.
\end{equation}
The results are given in \autoref{tab:comp-upper-Stahn}. The estimates are accurate (compared to the distance $d(\bE_{0}, [\octa])=6.49$ GPa and the relative distance $d(\bE_{0}, [\octa])/\norm{\bE_{0}}=0.221$ computed in~\cite{Fra1995}), and they are close to each other, meaning that the second-order covariants of $\bE_{0}$ carry here accurate information on the material symmetry coordinate system. We next provide the details concerning $\Delta_{\bb'}(\bE_{0}, [\octa])$, obtained in this case thanks to the minimization of
$\bb' : \bB_{\bH_{0}}\big|_{\HH^{2}} : \bb'$ (see Appendix \ref{subsec:Fra1995}). We get
\begin{equation*}
  \bb'=\left(
  \begin{array}{ccc}
    0.6282  & -0.1123 & 0.2526 \\
    -0.1123 & -0.6367 & 0.1534 \\
    0.2526  & 0.1534  & 0.0085 \\
  \end{array}
  \right)
  ,
\end{equation*}
and (in Kelvin notation):
\begin{equation*}
  [\bC_{\bb'}]=10^{-3}\,
  \left(
  \begin{array}{cccccc}
      -186           & 160           & 26             & -7\,\sqrt{2}   & -227\,\sqrt{2} & 37\,\sqrt{2}  \\
      160            & -226          & 66             & 186\,\sqrt{2}  & 48\,\sqrt{2}   & -98\,\sqrt{2} \\
      26             & 66            & -92            & -180\,\sqrt{2} & 180\,\sqrt{2}  & 61\,\sqrt{2}  \\
      -7\,\sqrt{2}   & 186\,\sqrt{2} & -180\,\sqrt{2} & 66 \cdot 2     & 61 \cdot 2     & 48 \cdot 2    \\
      -227\,\sqrt{2} & 48\,\sqrt{2}  & 180\,\sqrt{2}  & 61 \cdot 2     & 26 \cdot 2     & -7 \cdot 2    \\
      37\,\sqrt{2}   & -98\,\sqrt{2} & 61\,\sqrt{2}   & 48 \cdot 2     & -7 \cdot 2     & 160 \cdot 2   \\
    \end{array}
  \right),
\end{equation*}
so that
\begin{equation*}
  \bC_{\bb'}::\bH_{0}= -1.916
  \;\text{ GPa}.
\end{equation*}
The corresponding cubic elasticity tensor is (in Kelvin representation, and in the basis in which is expressed $[\bE_{0}]$)
\begin{equation*}
  [\bE]=\left(
  \begin{array}{cccccc}
      12.47            & 1.02             & 1.28             & 0.01 \,\sqrt{2}  & 0.44 \,\sqrt{2}  & -0.07 \,\sqrt{2} \\
      1.02             & 12.55            & 1.2              & -0.36 \,\sqrt{2} & -0.09 \,\sqrt{2} & 0.19 \,\sqrt{2}  \\
      1.28             & 1.2              & 12.29            & 0.34 \,\sqrt{2}  & -0.34 \,\sqrt{2} & -0.12 \,\sqrt{2} \\
      0.01 \,\sqrt{2}  & -0.36 \,\sqrt{2} & 0.34 \,\sqrt{2}  & 5.27 \cdot 2     & -0.12 \cdot 2    & -0.09 \cdot 2    \\
      0.44 \,\sqrt{2}  & -0.09 \,\sqrt{2} & -0.34 \,\sqrt{2} & -0.12 \cdot 2    & 5.34 \cdot 2     & 0.01 \cdot 2     \\
      -0.07 \,\sqrt{2} & 0.19 \,\sqrt{2}  & -0.12 \,\sqrt{2} & -0.09 \cdot 2    & 0.01 \cdot 2     & 5.09 \cdot 2     \\
    \end{array}
  \right)
  \;\text{ GPa}.
\end{equation*}

\begin{table}[h]
  \begin{center}
    \setlength{\arraycolsep}{1pt}
    \begin{tabular}{c|ccccc|c}
      \toprule
      $d(\bE_{0}, [\octa])$ &                    & $M=\Delta_{\bt_{0}}$ & $ \Delta_{\bd_{20}}$ & $ \Delta_{\ba'}$ & $ \Delta_{\bb'}$ & $\Delta_{\text{opt}}= \Delta_{\bt_{0}}$ \\
      \midrule
      6.49                  & Estimate (GPa):    & 7.809                & 7.818                & 7.848            & 7.621            & 7.809                                   \\
      \midrule
      0.221                 & Relative estimate: & 0.2660               & 0.2664               & 0.2674           & 0.2596           & 0.2660                                  \\
      \bottomrule
    \end{tabular}
    \caption{Comparison of upper bounds estimates of the distance to cubic elasticity $d(\bE_{0}, [\octa])$ for Vosges sandstone.}
    \label{tab:comp-upper-Francois}
  \end{center}
\end{table}

\section{Upper bounds estimates of the distance to orthotropy}
\label{sec:Upper bound-Ortho}

We consider still a given (measured) elasticity tensor  $\bE_{0}$, triclinic, with harmonic decomposition
\begin{equation*}
  \bE_{0}=(\lambda_{0}, \mu_{0}, \bd_{0}', \bv_{0}', \bH_{0}).
\end{equation*}
Its likely orthotropic coordinate system can be determined exactly as in the previous section, meaning that we assume
that it is one of the likely cubic coordinate systems, determined from either a second order-covariant of $\bE_{0}$ (among $\bt_{0}=\frac{2}{3}(\bd_{0}-\bv_{0})$ or $\bd_{20}=\bH_{0}\3dots \bH_{0}$, for instance), or from a deviatoric second-order tensor $\ba'$ or $\bb'$ which minimizes respectively~\eqref{eq:minTrHxa2} or~\eqref{eq:minB0b}. A proof that $\ba'$ carries information on the likely orthotropic basis has been given in section \ref{subsec:Likely-Ortho-Proof}. No such proof exists for the deviatoric tensor $\bb'$, which minimizes~\eqref{eq:minB0b}, but nothing prevents us from keeping it in the list $\set{\bt_{0}, \bd_{20}, \ba', \bb'}$ for comparison.

Once an orthotropic second order tensor that carries the likely orthotropic coordinate system has been exhibited, say $\ba$, with an orthotropic (dihedral) symmetry group $G_{\ba}$ (conjugate to $\DD_{2}$), we have to perform the orthogonal projection
\begin{equation*}
  \bE=\bR_{G_{\ba}}(\bE_0) =(\lambda_{0}, \mu_{0}, \bR_{G_{\ba}}(\bd_{0}'), \bR_{G_{\ba}}(\bv_{0}'), \bR_{G_{\ba}}(\bH_{0}))
\end{equation*}
onto the linear subspace $\Fix(G_{\ba})$ of orthotropic elasticity tensors, fixed by the group $G_{\ba}$ (see \autoref{subsec:SymClasses} and remark \ref{rem:RGEharm}). Instead of computing this projection using the Reynolds operator (by~\eqref{eq:D2}--\eqref{eq:ReyG}), we prefer to project each harmonic component of $\bE_{0}$, setting
\begin{equation*}
  \bd'=\bR_{G_{\ba}}(\bd_{0}'),
  \qquad
  \bv'=\bR_{G_{\ba}}(\bv_{0}'),
  \qquad
  \bH=\bR_{G_{\ba}}(\bH_{0}).
\end{equation*}
The orthotropic elasticity tensor $\bE$ that allows to define an upper bound estimate
\begin{equation}\label{eq:UpperOrtho}
  \Delta_{\ba}(\bE_{0}, [\DD_{2}])=\norm{\bE_{0}-\bE}
\end{equation}
of the distance $d(\bE_{0}, [\DD_{2}])$ to orthotropy is then, using~\eqref{eq:DecompHarm}
\begin{equation}\label{eq:EOrtho}
  \bE =  2 \mu_{0} \bI + \lambda_{0} \Id \otimes \Id + \frac{2}{7}\Id\odot (\bd'+2 \bv')+\Id\otimes_{(2,2)} (\bd'- \bv') + \bH.
\end{equation}

\begin{rem}
  Using deviatoric tensors, such as $\bt_{0}'$, $\bd_{20}'$, $\ba'$, $\bb'$, simplifies the orthogonal projection formulas derived thereafter for the fourth-order harmonic component $\bH$. However, even if an orthotropic second-order tensor $\ba$ is already deviatoric, we prefer to write $\Delta_{\ba'}(\bE_{0}, [\DD_{2}])$ instead of $\Delta_{\ba}(\bE_{0}, [\DD_{2}])$ for the sake of clarity (but we will keep the lighter notation $\ba$ in the calculations).
\end{rem}

\begin{rem}
  The Stahn and coworkers \emph{orthotropic upper bound estimate} $M(\bE_{0}, [\DD_{2}])$~\cite{SMB2020} is in fact equal to the upper bound estimate $\Delta_{\bt_{0}'}(\bE_{0}, [\DD_{2}])$ computed using $\ba=\bt_{0}'$ (see the results in Tables \ref{tab:comp-upper-ortho}, \ref{tab:comp-upper-Stahn-ortho} and \ref{tab:comp-upper-Francois-ortho}).
\end{rem}

The normal form \eqref{eq:VoigtOrtho} for the orthotropic estimate $\bE$ used to compute $\Delta_{\ba'}(\bE_{0}, [\DD_{2}])$ is recovered in any basis in which $\ba'$ is diagonal~\cite{Bae1993}.

\subsection{Orthogonal projection of harmonic components}
\label{subsec:Reynolds-projections}

Assume that the deviatoric tensor $\ba=\ba'$ is either a  second-order covariant of $\bE_{0}$ ($\bt_{0}'$ or $\bd_{20}'$ for example), or is the solution of either~\eqref{eq:minTrHxa2} or~\eqref{eq:minB0b}. Generically, $\ba$ is orthotropic and its symmetry group---a conjugate of $\DD_{2}$---is denoted by $G_{\ba}$.

The orthogonal projections $\bR_{G_{\ba}}(\bd_{0}')$ and $\bR_{G_{\ba}}(\bv_{0}')$ are obtained in an intrinsic way, as
\begin{equation}\label{eq:RGdv}
  \bd'=\bR_{G_{\ba}}(\bd_{0}')=\bP_{\ba}:\bd_{0}'
  ,\qquad
  \bv'=\bR_{G_{\ba}}(\bv_{0}')=\bP_{\ba}:\bv_{0}'
\end{equation}
using the following projector (a fourth-order rational covariant of $\ba$, introduced in~\cite{OD2021})
\begin{equation*}
  \bP_{\ba}=\bI-3 \frac{\left( \ba^{\, 2} \times \ba\right)\cdot \left( \ba^{\, 2} \times \ba\right)}{\norm{\ba^{\, 2} \times \ba}^2},
  \qquad \bI=\Id \otimesbar \Id.
\end{equation*}
Indeed, by~\eqref{eq:RelationPa}, it can be easily checked that in a basis $(\ee_{i})$ in which $\ba$ is diagonal, then, for any second-order tensor $\bb=(b_{ij})$ possibly non diagonal, we have
\begin{equation*}
  \bP_{\ba}:\bb =
  \begin{pmatrix} b_{11} & 0 & 0 \\
  0 & b_{22} &  0 \\
  0 & 0 &  b_{33}
  \end{pmatrix}
   = \bR_{G_{\ba}}(\bb).
\end{equation*}

Calculating an intrinsic expression of the orthogonal projection $\bR_{G_{\ba}}(\bH_{0})$ for $\bH_{0}\in \HH^{4}$ is less simple. It uses the fact that, by the theory of representations for tensor functions~\cite{Cow1985,Boe1987,Zhe1994,BB2001}, if $\ba$ is orthotropic, then any orthotropic harmonic fourth-order tensor $\bH\in \HH^{4}$ having the (orthotropic) symmetry group $G_{\ba}$, can be expressed as
\begin{equation*}
  \bH= \alpha\, \ba * \ba + \beta\, \ba * \ba^{2 \,\prime} + \gamma\, \ba^{2 \,\prime} * \ba^{2 \,\prime},
\end{equation*}
where the tensorial operation $\bh * \bk\in \HH^{4}$ is the \emph{harmonic product} (defined in~\cite{OKDD2018}) of two harmonic second-order tensors $\bh$, $\bk$ in $\HH^{2}$
\begin{align*}
  \bh * \bk:= (\bh \odot \bk)' & =\bh \odot \bk- \frac{2}{7} \, \Id \odot (\bh\bk+\bk\bh)'- \frac{2}{15}\,(\bh:\bk) \,  \Id \odot \Id,
  \\
                               & =\bh \odot \bk- \frac{2}{7}\, \Id \odot (\bh\bk+\bk\bh)+\frac{2}{35}\, (\bh:\bk) \,  \Id \odot \Id.
\end{align*}
The scalars $\alpha, \beta, \gamma$ are solution of the following system (since $\bH_{0}$ is harmonic)
\begin{equation*}
  [\,{\bf\mathcal G}\,]
  \begin{pmatrix}
    \alpha \\ \beta \\ \gamma
  \end{pmatrix}
  =
  \begin{pmatrix}
    \bH_{0}::(\ba * \ba) \\ \bH_{0}::(\ba * \ba^{2\, \prime})  \\ \bH_{0}::(\ba^{2\, \prime} * \ba^{2\, \prime})
  \end{pmatrix}
  =
  \begin{pmatrix}
    \ba:\bH_{0}:\ba \\ \ba:\bH_{0}:\ba^{2}  \\ \ba^{2}:\bH_{0}:\ba^{2}
  \end{pmatrix}
  ,
\end{equation*}
where $[\,\mathcal{G}\,]$ is the Gram matrix
\begin{equation*}
  [\,\mathcal{G}\,] =
  \begin{pmatrix}
    (\ba * \ba)::(\ba * \ba)                            & (\ba * \ba)::(\ba * \ba^{2\, \prime})                           & (\ba * \ba):(\ba^{2\, \prime} * \ba^{2\, \prime})                           \\
    (\ba * \ba^{2\, \prime})::(\ba * \ba)               & (\ba * \ba^{2\, \prime})::(\ba * \ba^{2\, \prime})              & (\ba * \ba^{2\, \prime}):(\ba^{2\, \prime} * \ba^{2\, \prime})              \\
    (\ba^{2\, \prime} * \ba^{2\, \prime}):: (\ba * \ba) & (\ba^{2\, \prime} * \ba^{2\, \prime})::(\ba * \ba^{2\, \prime}) & (\ba^{2\, \prime} * \ba^{2\, \prime}):(\ba^{2\, \prime} * \ba^{2\, \prime})
  \end{pmatrix}.
\end{equation*}
Nicely, Invariant Theory allows to express it as a function of the two fundamental invariants (here $I_{1}:=\tr \ba^{\prime} =0$)
\begin{equation*}
	I_{2}:=\tr(\ba^{\prime\, 2})
	\quad \text{and} \quad I_{3}:=\tr (\ba^{\prime\, 3}) ,
\end{equation*}
and we get
\begin{equation*}
  [\,\mathcal{G}\,] = \frac{1}{140}
  \left(
  \begin{array}{ccc}
      72 I_{2}^2           & 72 I_{2} I_{3}       & 2 I_{2}^3+60 I_{3}^2 \\
      72 I_{2} I_{3}       & 5 I_{2}^3+42 I_{3}^2 & 12 I_{2}^2 I_{3}     \\
      2 I_{2}^3+60 I_{3}^2 & 12 I_{2}^2 I_{3}     & 2 I_{2}^4            \\
    \end{array}
  \right).
\end{equation*}
Its determinant is (see~\eqref{eq:Bif-a-isoT})
  \begin{equation*}
    \frac{1}{3920} (I_{2}^3-6 I_{3}^2)^2= \frac{3}{980}\norm{\ba^{\, 2} \times \ba}^2.
  \end{equation*}
  It does not vanish if $\ba$ is orthotropic and, then, the inverse of $[\,\mathcal{G}\,]$ is a rational expression of $I_{2}$ and  $I_{3}$.

Finally, the orthogonal projection $\bR_{G_{\ba}}$ of $\HH^{4}$ on $\Fix(G_{\ba})$ is given by
\begin{equation}\label{eq:RGH}
  \bH = \bR_{G_{\ba}}(\bH_{0}) = \alpha\, \ba * \ba + \beta\, \ba * \ba^{2 \,\prime} + \gamma\, \ba^{2 \,\prime} * \ba^{2 \,\prime}
\end{equation}
where
\begin{equation}\label{eq:alphabetagamma}
  \begin{pmatrix}
    \alpha \\ \beta \\ \gamma
  \end{pmatrix}
  = \frac{1}{\left(I_{2}^3-6 I_{3}^2\right)^2}
  \left(
  \begin{array}{ccc}
      2 I_{2}^4            & -24 I_{2}^2 I_{3}      & 84 I_{3}^2-2 I_{2}^3 \\
      -24 I_{2}^2 I_{3}    & 28 I_{2}^3+120 I_{3}^2 & -144 I_{2} I_{3}     \\
      84 I_{3}^2-2 I_{2}^3 & -144 I_{2} I_{3}       & 72 I_{2}^2           \\
    \end{array}
  \right)
  \begin{pmatrix}
    \ba:\bH_{0}:\ba \\ \ba:\bH_{0}:\ba^{2}  \\ \ba^{2}:\bH_{0}:\ba^{2}
  \end{pmatrix}.
\end{equation}
The upper bound estimate $\Delta_{\ba'}(\bE_{0}, [\DD_{2}])$ of $d(\bE_{0}, [\DD_{2}])$, corresponding to the choice of $\ba=\ba'$ as the deviatoric second-order tensor that carries the likely orthotropic coordinate system of the raw elasticity tensor $\bE_{0}$, is then gained by~\eqref{eq:UpperOrtho}--\eqref{eq:EOrtho}, with $\bd'$ and $\bv'$ determined by~\eqref{eq:RGdv} and $\bH$ by~\eqref{eq:RGH}--\eqref{eq:alphabetagamma}.

\subsection{Example of Ni-based superalloy}

Consider first the triclinic elasticity tensor $\bE_{0}$ of the single crystal superalloy~\eqref{eq:E0}, close to be cubic, measured by François and coworkers. The tensor $\ba'$ represents one of the deviatoric second-order  tensors in the set $\set{\bt_{0}', \bd_{20}', \ba', \bb'}$, determined in section \ref{subsec:E0MF-cubic}. The upper bound estimates~\eqref{eq:UpperOrtho} as well as the corresponding relative estimates
\begin{equation*}
  \frac{M(\bE_{0}, [\DD_{2}])}{\norm{\bE_{0}}}=\frac{\Delta_{\bt_{0}'}(\bE_{0}, [\DD_{2}])}{\norm{\bE_{0}}}, \quad
  \frac{\Delta_{\bd_{20}'}(\bE_{0}, [\DD_{2}])}{\norm{\bE_{0}}}, \quad
  \frac{\Delta_{\ba'}(\bE_{0}, [\DD_{2}])}{\norm{\bE_{0}}}, \quad
  \frac{\Delta_{\bb'}(\bE_{0}, [\DD_{2}])}{\norm{\bE_{0}}},
\end{equation*}
are compared  in \autoref{tab:comp-upper-ortho}, respectively to the exact distance to orthotropy $d(\bE_{0}, [\DD_{2}])=57.8$ GPa and to the relative distance $d(\bE_{0}, [\octa])/\norm{\bE_0}=0.081$ (computed in~\cite{FGB1998}). Since the material has a cubic microstructure, it is not surprising that the second-order covariants do not carry much information on the likely orthotropic coordinate system (neither $\bt_{0}'$ nor $\bd_{20}'$), and that the orthotropic upper bounds estimates built from the fourth-order harmonic component $\bH_{0}$ are accurate. Surprisingly, the estimate $\Delta_{\bb'}(\bE_{0}, [\DD_{2}])$ is the more accurate. This example shows the interest of keeping $\bb'$ (which minimizes~\eqref{eq:minB0b}) in the list of second-order tensors which may carry the likely orthotropic coordinate system.

\begin{table}[h]
  \begin{center}
    \setlength{\arraycolsep}{1pt}
    \begin{tabular}{c|ccccc|c}
      \toprule
      $d(\bE_{0}, [\DD_{2}])$ &                    & $M=\Delta_{\bt_{0}'}$ & $ \Delta_{\bd_{20}'}$ & $ \Delta_{\ba'}$ & $ \Delta_{\bb'}$ & $\Delta_{\text{opt}}= \Delta_{\bb'}$
      \\
      \midrule
      57.8                    & Estimate (GPa):    & 216.1                 & 210.9                 & 109.8            & 90.3             & 90.3                                 \\
      \midrule
      0.081                   & Relative estimate: &
      0.3029                  & 0.2943             & 0.1539                & 0.1266                & 0.1266                                                                     \\
      \bottomrule
    \end{tabular}
    \caption{Comparison of upper bounds estimates of the distance to orthotropic elasticity $d(\bE_{0}, [\DD_{2}])$ for Ni-based single crystal superalloy.}
    \label{tab:comp-upper-ortho}
  \end{center}
\end{table}

The orthotropic elasticity tensor corresponding to
\begin{equation*}
  \Delta_{\bb'}(\bE_{0}, [\DD_{2}])=90.3 \quad \text{GPa}, \qquad \Delta_{\bb'}(\bE_{0}, [\DD_{2}])/\norm{\bE_{0}}=0.127,
\end{equation*}
is (in Kelvin representation, and in the basis in which is expressed $[\bE_{0}]$):
\begin{equation*}
  [\bE]=\left(
  \begin{array}{cccccc}
      239.12           & 136.25            & 131.67            & -0.44 \, \sqrt{2}  & 48.41\, \sqrt{2}  & -1.36 \, \sqrt{2} \\
      136.25           & 236.12            & 143.68            & -16.16 \, \sqrt{2} & 2.00 \, \sqrt{2}  & -3.38\, \sqrt{2}  \\
      131.67           & 143.68            & 232.58            & 17.32\, \sqrt{2}   & -50.18\, \sqrt{2} & 4.89\, \sqrt{2}   \\
      -0.44\, \sqrt{2} & -16.16\, \sqrt{2} & 17.32\, \sqrt{2}  & 133.07 \, \cdot 2  & 5.05 \, \cdot 2   & 0.34 \, \cdot 2   \\
      48.41\, \sqrt{2} & 2.00\, \sqrt{2}   & -50.18\, \sqrt{2} & 5.05 \, \cdot 2    & 120.55 \, \cdot 2 & 0.20 \, \cdot 2   \\
      -1.36\, \sqrt{2} & -3.38\, \sqrt{2}  & 4.89\, \sqrt{2}   & 0.34 \, \cdot 2    & 0.20 \, \cdot 2   & 131.97 \, \cdot 2 \\
    \end{array}
  \right)
  \;\text{ GPa}.
\end{equation*}
It has been computed by using the symmetry group $G_{\bb}$ of the second-order tensor $\bb$ given by~\eqref{eq:abFGB1998}.

\subsection{Academic example of a close to be orthotropic material}
\label{subsec:SMB-ortho}

For the close to be orthotropic elasticity tensor~\eqref{eq:E0Stahn}, and the same set of second-order tensors $\set{\bt_{0}', \bd_{20}', \ba', \bb'}$ defined in \autoref{subsec:E0SMB-cubic}, we find that the best estimate is the Stahn and coworkers upper bound estimate $M(\bE_{0}, [\DD_{2}])=\Delta_{\bt_{0}'}(\bE_{0}, [\DD_{2}])$ of the distance to orthotropy. All the estimates are nevertheless quite good: in this case both the second-order covariants $\bt_{0}'$, $\bd_{20}'$ as well as the fourth-order covariant $\bH_{0}$ of $\bE_{0}$ carry precise informations on the likely orthotropic coordinate system.

\begin{table}[h!]
  \begin{center}
    \setlength{\arraycolsep}{1pt}
    \begin{tabular}{ccccc|c}
      \toprule
                         & $M=\Delta_{\bt_{0}'}$                     & $ \Delta_{\bd_{20}'}$ &
      $ \Delta_{\ba'}$   &
      $ \Delta_{\bb'}$   & $ \Delta_{\text{opt}}= \Delta_{\bt_{0}'}$
      \\
      \midrule
      Estimate:          & 0.491                                     & 0.651                 & 0.653  & 6.714  & 0.491
      \\
      \midrule
      Relative estimate: & 0.0175                                    & 0.02325               & 0.0233 & 0.2398 & 0.01753
      \\
      \bottomrule
    \end{tabular}
    \caption{Comparison of upper bounds estimates of the distance to orthotropic elasticity $d(\bE_{0}, [\DD_{2}])$ for an academic elasticity tensor.}
    \label{tab:comp-upper-Stahn-ortho}
  \end{center}
\end{table}

The orthotropic elasticity tensor corresponding to $M(\bE_{0}, [\DD_{2}])=\Delta_{\bt_{0}'}(\bE_{0}, [\DD_{2}])=0.491$ GPa,
$M(\bE_{0}, [\DD_{2}])/\norm{\bE_{0}}=0.0175$, is (in Kelvin representation, and in the basis in which is expressed $[\bE_{0}]$)
\begin{equation*}
  [\bE]=\left(
  \begin{array}{cccccc}
      9.28  & 5.85  & -0.17 & 5.01  & 4.03  & -2.71 \\
      5.85  & 11.13 & 2.61  & -0.85 & -3.94 & 2.67  \\
      -0.17 & 2.61  & 11.02 & -0.11 & -3.08 & 0.54  \\
      5.01  & -0.85 & -0.11 & 8.02  & -1.20 & 0.77  \\
      4.03  & -3.94 & -3.08 & -1.20 & 8.00  & 1.88  \\
      -2.71 & 2.67  & 0.54  & 0.77  & 1.88  & 8.20  \\
    \end{array}
  \right)
  \;\text{ GPa}.
\end{equation*}
It has been computed using the symmetry group $G_{\bt_0'}$ of the second-order tensor
\begin{equation*}
  \bt_{0}'= \left(
  \begin{array}{ccc}
      0.0811  & 0.6554 & -2.1146 \\
      0.6554  & 1.9311 & 1.7342  \\
      -2.1146 & 1.7342 & -2.0122 \\
    \end{array}
  \right)
  \;\text{ GPa}.
\end{equation*}

\subsection{Example of Vosges sandstone}

Consider finally the raw elasticity tensor~\eqref{eq:E0Francois} identified by François, also not so far from being orthotropic (with a distance $d(\bE_{0}, [\DD_{2}])=2.64$ GPa and a relative distance to orthotropy $d(\bE_{0}, [\DD_{2}])/\norm{\bE_{0}}=0.090$ computed in~\cite{Fra1995}). The set of second-order tensors $\set{\bt_{0}', \bd_{20}', \ba', \bb'}$ is the one computed in section
\ref{subsec:E0Francois-cubic}. The four estimates are similarly good (with relative distance $\approx 0.15$), the the best of them being the one obtained by Stahn and coworkers $M(\bE_{0}, [\DD_{2}])=\Delta_{\bt_{0}'}(\bE_{0}, [\DD_{2}])=0.149$.

\begin{table}[h!]
  \begin{center}
    \setlength{\arraycolsep}{1pt}
    \begin{tabular}{c|ccccc|c}
      \toprule
      $d(\bE_{0}, [\DD_{2}])$ &                    & $M=\Delta_{\bt_{0}'}$ & $ \Delta_{\bd_{20}'}$ & $ \Delta_{\ba'}$ & $ \Delta_{\bb'}$ & $\Delta_{\text{opt}}= \Delta_{\bt_{0}'}$
      \\
      \midrule
      2.64                    & Estimate (GPa):    & 4.3756                & 4.5587                & 4.8292           & 4.5584           & 4.3756                                   \\
      \midrule
      0.090                   & Relative estimate: & 0.14907               & 0.15531               & 0.16453          & 0.15530          & 0.14907                                  \\
      \bottomrule
    \end{tabular}
    \caption{Comparison of upper bounds estimates of the distance to orthotropic elasticity $d(\bE_{0}, [\DD_{2}])$ for Vosges sandstone.}
    \label{tab:comp-upper-Francois-ortho}
  \end{center}
\end{table}

The orthotropic elasticity tensor corresponding to $M(\bE_{0}, [\DD_{2}])=\Delta_{\bt_{0}'}(\bE_{0}, [\DD_{2}])=4.38$ GPa
and to a relative distance to orthotropy of $0.149$ is (in Kelvin representation, and in the basis in which is expressed $[\bE_{0}]$)
\begin{equation*}
  [\bE]=\left(
  \begin{array}{cccccc}
      11.35            & -1.20            & 2.18             & 0.65\, \sqrt{2} & -0.05\, \sqrt{2} & -0.55\, \sqrt{2} \\
      -1.20            & 11.52            & 3.15             & 0.02\, \sqrt{2} & -0.15\, \sqrt{2} & 0.48\, \sqrt{2}  \\
      2.18             & 3.15             & 13.17            & 0.51\, \sqrt{2} & -0.21\, \sqrt{2} & -0.44\, \sqrt{2} \\
      0.65\, \sqrt{2}  & 0.02\, \sqrt{2}  & 0.51\, \sqrt{2}  & 4.25 \cdot 2    & 0.66 \cdot 2     & 0.15 \cdot 2     \\
      -0.05\, \sqrt{2} & -0.15\, \sqrt{2} & -0.21\, \sqrt{2} & 0.66 \cdot 2    & 5.68 \cdot 2     & -0.25 \cdot 2    \\
      -0.55\, \sqrt{2} & 0.48\, \sqrt{2}  & -0.44\, \sqrt{2} & 0.15 \cdot 2    & -0.25 \cdot 2    & 6.40 \cdot 2     \\
    \end{array}
  \right)
  \;\text{ GPa}.
\end{equation*}
It has been computed using the symmetry group $G_{\bt_0'}$ of the second-order tensor
\begin{equation*}
  \bt_{0}'=\left(
  \begin{array}{ccc}
      -1.9778 & -0.7333 & -0.2000 \\
      -0.7333 & -0.3778 & 0.6000  \\
      -0.2000 & 0.6000  & 2.3556  \\
    \end{array}
  \right)
  \;\text{ GPa}.
\end{equation*}

\section{Closure}

We have addressed the problem not of the determination---which usually needs numerical methods---but of an accurate analytical estimation of the distance of a raw elasticity tensor  $\bE_{0}$ either to cubic symmetry or to orthotropy. Following~\cite{Bae1993,SMB2020}, the key point of the present work is the use of a second-order tensor $\ba$ built from $\bE_{0}$ and which carries the likely symmetry coordinate system. The Stahn and coworker upper bounds estimates $M(\bE_{0}, G)$ \cite{SMB2020} of both the distance to cubic symmetry ($G=\octa$) and the distance to orthotropy ($G=\DD_{2}$), correspond both to the particular choice ($\ba=\bt_{0}'$) of a second-order covariant of $\bE_{0}$ used to compute the more general estimate $\Delta_{\ba}(\bE_{0}, G)$. There are, however, not only one but many possible second-order covariants that can be used to build an upper bound estimate of the distance (see the list in remark~\ref{rem:many-cov2} and~\cite{OKDD2021,DADKO2019}). Furthermore, since all second-order covariants of a cubic elasticity tensor are isotropic, these estimates are not accurate for a material with a cubic microstructure (such as a Ni-based single crystal superalloy, see \autoref{tab:comp-upper}), even if its elasticity tensor $\bE_{0}$ is measured as triclinic.

This observation has led us to suggest another way to determine a deviatoric second-order tensor (denoted $\ba'$ or $\bb'$), which carries its likely cubic/orthotropic coordinate system. This tensor is not a covariant of $\bE_{0}$. It is obtained as the solution of a minimization problem (reducing to an eigenvalue problem of a symmetric operator), which is a generalization of an approach proposed by Klimeš for transverse isotropy~\cite{Kli2016,Kli2018}.

More precisely, we formulate a quadratic functional defined on symmetric and deviatoric second-order tensors (and depending on $\bE_{0}$). This functional does not vanish in general. But when $\bE_{0}$ is cubic or orthotropic, orthotropic deviators on which this functional vanishes are precisely those whose eigenvectors define a natural basis for $\bE_{0}$. It is therefore expected that, when $\bE_{0}$ is triclinic, but possibly not too far from a cubic or an orthotropic tensor, the minimizer of this functional will provide an approximation of the likely symmetry coordinate system.

Having introduced several manners to produce an orthotropic symmetric second-order tensor whose eigenvectors approximate accurately the likely symmetry coordinate system of an experimental (an expected cubic or orthotropic) tensor $\bE_{0}$, we are able to improve the calculation of an upper bound estimate of the distance of $\bE_{0}$ to the cubic or orthotropic symmetry (see Tables \ref{tab:comp-upper} to \ref{tab:comp-upper-Francois-ortho}, using not only one but several second-order tensors).

Finally, note that the optimal tensor $\bE$, used to define an upper bound estimate as $\norm{\bE_{0}-\bE}$, is determined a priori, \emph{i.e.}, before the calculation of the distance estimate in question. This allows to consider readily other norms than the Euclidean norm~\cite{MN2006,MGD2019}. The example of upper bounds estimates defined from the Log-Euclidean norm~\cite{AFPA2005} is provided in \autoref{sec:Log}.

\appendix

\section{Representative symmetry groups}
\label{sec:groups}

Some representative subgroups $G\subset \SO(3)$ are:
\begin{itemize}
  \item $\ZZ_{2}$, of order $\abs{\ZZ_{2}}=2$, generated by the second-order rotation $\rot(\ee_{3},\pi)$,
  \item $\DD_{2}$, of order $\abs{\DD_{2}}=4$, generated by the second-order rotations $\rot(\ee_{3},\pi)$ and $\rot(\ee_{1},\pi)$,
  \item $\DD_{3}$, of order $\abs{\DD_{3}}=6$, generated by the third order rotation $\rot(\ee_{3},\frac{2\pi}{3})$ and the second-order rotation $\rot(\ee_{1},\pi)$,
  \item $\DD_{4}$, of order $\abs{\DD_{2}}=8$, generated by the fourth-order rotation $\rot(\ee_{3},\frac{\pi}{2})$ and the second-order rotation $\rot(\ee_{1},\pi)$,
  \item $\tetra$, the \emph{tetrahedral} group, order $\abs{\tetra}=12$, which is the orientation-preserving symmetry group of the tetrahedron; it is generated by
          $\rot(\ee_{3},\pi )$, $\rot(\ee_{1},\pi )$ and $\left.\rot(\ee_{1}+\ee_{2}+\ee_{3},\frac{2\pi}{3})\right.$,
  \item $\octa$, the proper \emph{octahedral} group, of order $\abs{\octa}=24$, the orientation-preserving symmetry group of the cube with vertices $(\pm 1,\pm 1,\pm 1)$;
        its principal directions are the normals to its faces, which are the basis vectors $\pm \ee_{i}$,
  \item $\OO(2)$, of infinite order, generated by all rotations $\rot(\ee_{3},\theta)$ ($\theta\in [0;2\pi[$) and the second-order rotation $\rot(\ee_{1},\pi)$,
  \item $\SO(2)$, of infinite order, generated by all rotations $\rot(\ee_{3},\theta)$ ($\theta\in [0;2\pi[$).
\end{itemize}

\section{Stahn and coworkers upper bounds estimates}
\label{sec:Stahn}

Stahn and coworkers define in practice their upper bound estimate $M(\bE_{0}, [G])$ of $d(\bE_{0}, [G])$
by the following steps~\cite{SMB2020}:
\begin{enumerate}
  \item compute an eigenbasis of
        \begin{equation*}
          \bt_{0} =\Id:(\bE_0-\bE_0^s)=\frac{2}{3}(\bd_{0}-\bv_{0}),
        \end{equation*}
        and a rotation $r_{0}$ that brings it into its diagonal form
        (\emph{i.e.}, such that $r_{0}\star \bt_{0}=r_{0} \, \bt_{0}\, r_{0}^{T}$ is diagonal),

  \item compute the rotated elasticity tensor $\bE_{r_{0}}=r_{0} \star \bE_{0}$,

  \item compute the subset
        \begin{equation*}
          S=\set{\bE_{n} =r_{n}^{T} \star \bR_{G}\left(r_{n}  \star \bE_{r_{0}}\right), \; r_{n}\in \octa} \subset \strata{G},
        \end{equation*}
        of $N$ elasticity tensors built from $\bE_{0}$ but of the symmetry class $[G]$.

  \item set
        \begin{equation}\label{eq:MStahn}
          M(\bE_{0}, [G]): = \min_{r \in \octa}
          \norm{r \star\bE_{r_{0}} - \bR_{G}\left(r \star \bE_{r_{0}}\right) }
          = \min_{1\leq n \leq N}
          \norm{\bE_{r_{0}} - \bE_{n} },
        \end{equation}
\end{enumerate}

The definition~\eqref{eq:MStahn} is independent of the choice of the rotation $r$, among the 24 possibilities, which brings $\bt_0$ into its diagonal form.

\section{Klimeš quadratic form and upper bound estimate of $d(\bE_{0}, [\OO(2)])$}
\label{sec:Klimes}

Let $\bE$ be an elasticity tensor, with harmonic decomposition
\begin{equation*}
  \bE = (\lambda, \mu, \bd', \bv', \bH ),
\end{equation*}
and let
\begin{equation*}
  \bE^s = (2 \mu+\lambda) \Id \odot \Id +\frac{2}{7} \Id \odot (\bd' +2 \bv' )+\bH
\end{equation*}
be its totally symmetry part. Klimeš tensor~\eqref{eq:KlimesT} can be recast as
\begin{align*}
  \bT & =16 \left(\bE^s \times \nn + \Id \otimes_{(2,2)} \left[\left(\bd'-\bv'\right)\times \nn\right]\right)
  \\ &=16 \left(\bH \times \nn + \frac{1}{7} \Id \odot \left[\left(\bd'+2\bv'\right)\times \nn\right]+ \Id \otimes_{(2,2)} \left[\left(\bd'-\bv'\right)\times \nn\right]\right).
\end{align*}
In particular, the harmonic decomposition of $\bT$ is given by
\begin{equation*}
  \bT = (0, 0,  8\, \bd'\times \nn, 8\, \bv'\times \nn, 16\, \bH \times \nn).
\end{equation*}
By remark~\ref{rem:normE2}, we get thus
\begin{equation}
  \norm{\bT}^{2} = 64\left(\frac{2}{21} \norm{(\bd^{\prime}+2 \bv^{\prime})\times \nn}^{2}
  +\frac{4}{3}\norm{(\bd^{\prime}- \bv^{\prime})\times \nn}^{2}+ 4\norm{\bH\times \nn}^{2}\right),
\end{equation}
so that
\begin{align*}
  \norm{\bT}^{2} = 64 \Big( & \frac{2}{21} \norm{\bd^{\prime}+2 \bv^{\prime}}^{2}\norm{\nn}^{2} -\frac{1}{7} \nn\cdot(\bd^{\prime}+2 \bv^{\prime})^{2}\cdot \nn
  +
  \frac{4}{3} \norm{\bd^{\prime}- \bv^{\prime}}^{2}\norm{\nn}^{2}- 2 \nn\cdot(\bd^{\prime}- \bv^{\prime})^{2}\cdot \nn
  \\
                            & + 4 \norm{\bH}^{2} \norm{\nn}^{2}- 7\, \nn \cdot \bd_{2} \cdot \nn \Big),
\end{align*}
and finally
\begin{equation*}
  \norm{\bT}^{2} = \nn\cdot A(\bE)\cdot \nn,
\end{equation*}
where
\begin{align*}
  A(\bE)= 64 & \left[ \left(4 \,\norm{\bH}^{2}+ \frac{2}{21} \norm{\bd^{\prime}+2 \bv^{\prime}}^{2}+ \frac{4}{3} \norm{\bd^{\prime}- \bv^{\prime}}^{2}\right) \Id
    -7\, \bd_{2} -\frac{1}{7} (\bd^{\prime}+2\, \bv^{\prime})^{2}-2\, (\bd^{\prime}- \bv^{\prime})^{2}
    \right].
\end{align*}

The solution of the problem
\begin{equation*}
  \min_{\norm{\nn}=1} \nn\cdot A(\bE)\cdot \nn
\end{equation*}
is the unit eigenvector $\nn$ of $A$ corresponding to its smallest eigenvalue. Once such a unit vector $\nn$ is known, one can build
  a transversely isotropic deviatoric tensor
  \begin{equation*}
    \bt :=  \nn \ast \nn =  (\nn \otimes \nn)',
    \qquad
    \norm{\bt}= \sqrt{\frac{2}{3}}.
  \end{equation*}
  with transverse isotropy axis $\langle \nn \rangle$ and symmetry group $G_{\bt}$. Using the formulas in~\cite[Appendix C]{DAD2022}, Klimeš upper bound estimate $K(\bE_{0},  \strata{\OO(2)})$~\cite{Kli2018}, where $\bE_{0}=(\lambda_{0},\, \mu_{0}, \,\bd_{0}', \bv_{0}',\, \bH_{0})$ recast as
\begin{equation*}
  K(\bE_{0}, \strata{\OO(2)})=\norm{\bE_{0}-\bE},
\end{equation*}
where $\bE$ is the orthogonal projection of $\bE_{0}$ onto $\Fix(G_{\bt})$, which is given by
\begin{equation}\label{eq:EGt}
  \bE=\left(\lambda_{0},\; \mu_{0}, \;\bd'=\frac{3}{2} (\bd_{0}':\bt) \,\bt, \bv'=\frac{3}{2} (\bv_{0}':\bt) \,\bt,\; \bH=\frac{35}{8} (\bt:\bH_{0}:\bt)\, \bt \ast \bt\right),
\end{equation}
and
\begin{equation*}
  \bt \ast \bt = \bt \otimes \bt -\frac{4}{7} \Id \odot \bt^{2} +\frac{2}{35} \norm{\bt}^{2} \Id \odot \bt \in \HH^{4}.
\end{equation*}

\begin{rem}
  The harmonic components~\eqref{eq:EGt} of the estimated transversely isotropic elasticity tensor $\bE$ correspond indeed to the Reynolds averaging for the symmetry group $G=G_{\bt}$, of the harmonic components of $\bE_{0}$,
  \begin{align*}
     & \bd'=\bR_{G_{\bt}}(\bd_{0}')=\frac{3}{2} (\bd_{0}':\bt)\, \bt,
    \\
     & \bv'=\bR_{G_{\bt}}(\bv_{0}')=\frac{3}{2} (\bv_{0}':\bt)\, \bt,
    \\
     & \bH=\bR_{G_{\bt}}(\bH_{0})=\frac{35}{8} (\bt:\bH_{0}:\bt)\, \bt \ast \bt.
  \end{align*}
\end{rem}

\section{Components of third-order tensors $\tr(\bH \times \ba)$ and $\ba \times \bb$}
\label{sec:TrHxa}

When $\bH$ is a totally symmetric or an harmonic fourth-order tensor and $\ba$ is a symmetric second order tensor, the ten independent components of the totally symmetric third order tensor $\tr(\bH \times \ba)$ are:
{\footnotesize
\begin{equation*}
  \begin{aligned}
    (\tr(\bH \times \ba))_{111} = & \frac{3}{10} \Big(-a_{12} (2 H_{1113}+H_{1223}+H_{1333})+a_{13} (2 H_{1112}+H_{1222}+H_{1233})-a_{22} H_{1123}+a_{23} H_{1122}
    \\
                                  & -a_{23} H_{1133}+a_{33} H_{1123}\Big),
    \\
    (\tr(\bH \times \ba))_{112} = & \frac{1}{10} \Big(a_{11} (2 H_{1113}+H_{1223}+H_{1333})-2 a_{12} H_{1123}-a_{12} H_{2223}-a_{12} H_{2333}
    \\
                                  & +a_{13} (-2 H_{1111}+2 H_{1122}+H_{2222}+H_{2233})-a_{22} H_{1113}-3 a_{22} H_{1223}-a_{22} H_{1333}
    \\
                                  & +3 a_{23} H_{1222}-a_{23} H_{1233}-a_{33} H_{1113}+2 a_{33} H_{1223}\Big),
    \\
    (\tr(\bH \times \ba))_{113} = & \frac{1}{10} \Big(-a_{11} (2 H_{1112}+H_{1222}+H_{1233})+a_{12} (2 H_{1111}-2 H_{1133}-H_{2233}-H_{3333})+2 a_{13} H_{1123}
    \\
                                  & +a_{13} H_{2223}+a_{13} H_{2333}+a_{22} H_{1112}-2 a_{22} H_{1233}+a_{23} H_{1223}-3 a_{23} H_{1333}+a_{33} H_{1112}
    \\
                                  & +a_{33} H_{1222}+3 a_{33} H_{1233}\Big),
    \\
    (\tr(\bH \times \ba))_{122} = & \frac{1}{10} \Big(3 a_{11} H_{1123}+a_{11} H_{2223}+a_{11} H_{2333}+a_{12} H_{1113}+2 a_{12} H_{1223}+a_{12} H_{1333}+a_{13} (H_{1233}-3 H_{1112})
    \\
                                  & -a_{22} H_{1123}-2 a_{22} H_{2223}-a_{22} H_{2333}-a_{23} (H_{1111}+2 H_{1122}+H_{1133}-2 H_{2222})-2 a_{33} H_{1123}+a_{33} H_{2223}\Big),
    \\
    (\tr(\bH \times \ba))_{123} = & \frac{1}{20} \Big(-3 a_{11} H_{1122}+3 a_{11} H_{1133}-a_{11} H_{2222}+a_{11} H_{3333}+2 a_{12} H_{1112}-2 a_{12} H_{1222}-2 a_{13} H_{1113}
    \\
                                  & +2 a_{13} H_{1333}+a_{22} (H_{1111}+3 H_{1122}-3 H_{2233}-H_{3333})+2 a_{23} H_{2223}-2 a_{23} H_{2333}
    \\
                                  & +a_{33} (-H_{1111}-3 H_{1133}+H_{2222}+3 H_{2233})\Big),
    \\
    (\tr(\bH \times \ba))_{133} = & \frac{1}{10} \Big(-3 a_{11} H_{1123}-a_{11} H_{2223}-a_{11} H_{2333}+3 a_{12} H_{1113}-a_{12} H_{1223}-a_{13} (H_{1112}+H_{1222}+2 H_{1233})
    \\
                                  & +2 a_{22} H_{1123}-a_{22} H_{2333}+a_{23} (H_{1111}+H_{1122}+2 H_{1133}-2 H_{3333})+a_{33} H_{1123}+a_{33} H_{2223}+2 a_{33} H_{2333}\Big),
    \\
    (\tr(\bH \times \ba))_{222} = & \frac{3}{10} \Big(a_{11} H_{1223}+a_{12} H_{1123}+2 a_{12} H_{2223}+a_{12} H_{2333}+a_{13} (H_{2233}-H_{1122})
    \\
                                  & -a_{23} (H_{1112}+2 H_{1222}+H_{1233})-a_{33} H_{1223}\Big),
    \\
    (\tr(\bH \times \ba))_{223} = & \frac{1}{10} \Big(-a_{11} H_{1222}+2 a_{11} H_{1233}+a_{12} H_{1133}-2 a_{12} H_{2222}+2 a_{12} H_{2233}+a_{12} H_{3333}-a_{13} H_{1123}+3 a_{13} H_{2333}
    \\
                                  & +a_{22} (H_{1112}+2 H_{1222}+H_{1233})
    -a_{23} H_{1113}-2 a_{23} H_{1223}-a_{23} H_{1333}-a_{33} (H_{1112}+H_{1222}+3 H_{1233})\Big),
    \\
    (\tr(\bH \times \ba))_{233} = & \frac{1}{10} \Big(-2 a_{11} H_{1223}+a_{11} H_{1333}+a_{12} H_{1123}-3 a_{12} H_{2223}-a_{13} H_{1122}-a_{13} H_{2222}-2 a_{13} H_{2233}+2 a_{13} H_{3333}
    \\
                                  & +a_{22} (H_{1113}+3 H_{1223}+H_{1333})+a_{23} (H_{1112}+H_{1222}+2 H_{1233})-a_{33} H_{1113}-a_{33} H_{1223}-2 a_{33} H_{1333}\Big),
    \\
    (\tr(\bH \times \ba))_{333} = & \frac{3}{10} \Big(-a_{11} H_{1233}+a_{12} H_{1133}-a_{12} H_{2233}-a_{13} (H_{1123}+H_{2223}+2 H_{2333})+a_{22} H_{1233}
    \\
                                  & +a_{23} (H_{1113}+H_{1223}+2 H_{1333})\Big).
  \end{aligned}
\end{equation*}
}

The 10 independent components of the totally symmetric third order tensor $\ba \times \bb$, where both $\ba$ and $\bb$ are symmetric second order tensors, are:
\begin{equation*}
  \begin{aligned}
     & (\ba \times \bb)_{111} =a_{12} b_{13}-a_{13} b_{12},
    \\
     & (\ba \times \bb)_{112} = \frac{1}{3} (-a_{11} b_{13}+a_{12} b_{23}+a_{13} b_{11}-a_{13} b_{22}+a_{22} b_{13}-a_{23} b_{12}),
    \\
     & (\ba \times \bb)_{113} =\frac{1}{3} (a_{11} b_{12}-a_{12} b_{11}+a_{12} b_{33}-a_{13} b_{23}+a_{23} b_{13}-a_{33} b_{12}),
    \\
     & (\ba \times \bb)_{122} = \frac{1}{3} (-a_{11} b_{23}-a_{12} b_{13}+a_{13} b_{12}+a_{22} b_{23}+a_{23} b_{11}-a_{23} b_{22}),
    \\
     & (\ba \times \bb)_{123} =  \frac{1}{6} (a_{11} b_{22}- a_{11}  b_{33}+a_{22} b_{33}-a_{22}b_{11}+a_{33} b_{11}-a_{33}b_{22}),
    \\
     & (\ba \times \bb)_{133} = \frac{1}{3} (a_{11} b_{23}-a_{12} b_{13}+a_{13} b_{12}-a_{23} b_{11}+a_{23} b_{33}-a_{33} b_{23}),
    \\
     & (\ba \times \bb)_{222} =a_{23} b_{12}-a_{12} b_{23},
    \\
     & (\ba \times \bb)_{223} =\frac{1}{3} (a_{12} b_{22}-a_{12} b_{33}-a_{13} b_{23}-a_{22} b_{12}+a_{23} b_{13}+a_{33} b_{12}),
    \\
     & (\ba \times \bb)_{233} = \frac{1}{3} (a_{12} b_{23}+a_{13} b_{22}- a_{13} b_{33}-a_{22} b_{13}-a_{23} b_{12}+a_{33} b_{13}),
    \\
     & (\ba \times \bb)_{333} =a_{13} b_{23}-a_{23} b_{13}.
  \end{aligned}
\end{equation*}

\section{Raw elasticity tensors and their harmonic components}

\subsection{Elasticity tensor of François--Geymonat--Berthaud~\cite{FGB1998} for Ni-based single crystal superalloy}
\label{subsec:FGB1998}

The harmonic components of the elasticity tensor~\eqref{eq:E0} are:
\begin{equation*}
  \lambda_{0}=\frac{1583}{15}=105.533\;  \text{ GPa},
  \qquad
  \mu_{0}=\frac{1453}{15}=96.867\;  \text{ GPa},
\end{equation*}
\begin{equation*}
  \bd_{0}'=\left(
  \begin{array}{ccc}
      \frac{11}{3} & 2           & 14            \\
      2            & \frac{5}{3} & 23            \\
      14           & 23          & -\frac{16}{3} \\
    \end{array}
  \right) \text{ GPa},
  \qquad
  \bv_{0}'=\left(
  \begin{array}{ccc}
      -1  & -11 & -1 \\
      -11 & 9   & -1 \\
      -1  & -1  & -8 \\
    \end{array}
  \right)
  \text{ GPa},
\end{equation*}
and, by~\eqref{eq:H} (in Kelvin notation),
\begin{equation*}
  [\bH_{0}]=\frac{1}{35}\left(
  \begin{array}{cccccc}
      -1986           & 1093             & 893              & 175 \,\sqrt{2}   & 1760 \,\sqrt{2}  & -495 \,\sqrt{2} \\
      1093            & -2306            & 1213             & -1085 \,\sqrt{2} & 15 \,\sqrt{2}    & 660 \,\sqrt{2}  \\
      893             & 1213             & -2106            & 910 \,\sqrt{2}   & -1775 \,\sqrt{2} & -165 \,\sqrt{2} \\
      175 \,\sqrt{2}  & -1085 \,\sqrt{2} & 910 \,\sqrt{2}   & 1213 \cdot 2     & -165 \cdot 2     & 15 \cdot 2      \\
      1760 \,\sqrt{2} & 15 \,\sqrt{2}    & -1775 \,\sqrt{2} & -165 \cdot 2     & 893 \cdot 2      & 175 \cdot 2     \\
      -495 \,\sqrt{2} & 660 \,\sqrt{2}   & -165 \,\sqrt{2}  & 15 \cdot 2       & 175 \cdot 2      & 1093 \cdot 2    \\
    \end{array}
  \right)\text{ GPa},
\end{equation*}
so that
\begin{equation*}
  \bd_{20}'=
  \left(
  \begin{array}{ccc}
      133.497  & -1440.53 & -1055.76 \\
      -1440.53 & -3457.12 & 827.469  \\
      -1055.76 & 827.469  & 3323.62  \\
    \end{array}
  \right)
  \text{ GPa}^{2}.
\end{equation*}

Finally (in Kelvin notation):
\begin{equation*}
  [\bA_{0}]=
  \left(
  \begin{array}{cccccc}
      195.6             & -7.3              & -188.3            & 34.7 \,\sqrt{2}   & 216. \,\sqrt{2}   & -199.1 \,\sqrt{2} \\
      -7.3              & 52.8              & -45.5             & -107.2 \,\sqrt{2} & 51.9 \,\sqrt{2}   & 39. \,\sqrt{2}    \\
      -188.3            & -45.5             & 233.8             & 72.5 \,\sqrt{2}   & -267.9 \,\sqrt{2} & 160.1 \,\sqrt{2}  \\
      34.7 \,\sqrt{2}   & -107.2 \,\sqrt{2} & 72.5 \,\sqrt{2}   & 677.1 \cdot 2     & 149.3\cdot 2      & 44. \cdot 2       \\
      216.  \,\sqrt{2}  & 51.9 \,\sqrt{2}   & -267.9 \,\sqrt{2} & 149.3 \cdot 2     & 561.2 \cdot 2     & 40.9 \cdot 2      \\
      -199.1 \,\sqrt{2} & 39.  \,\sqrt{2}   & 160.1 \,\sqrt{2}  & 44. \cdot 2       & 40.9 \cdot 2      & 691.3 \cdot 2     \\
    \end{array}
  \right)
  \text{ GPa}^{2},
\end{equation*}
\begin{equation*}
  [\bB_{0}]=\left(
  \begin{array}{cccccc}
      183.6             & -306.5            & 122.9             & 109.2 \,\sqrt{2}  & 263.5 \,\sqrt{2}  & -134.2 \,\sqrt{2} \\
      -306.5            & 364.              & -57.5             & -144.5 \,\sqrt{2} & -43.1 \,\sqrt{2}  & 103.8 \,\sqrt{2}  \\
      122.9             & -57.5             & -65.3             & 35.2 \,\sqrt{2}   & -220.4 \,\sqrt{2} & 30.4 \,\sqrt{2}   \\
      109.2 \,\sqrt{2}  & -144.5 \,\sqrt{2} & 35.2 \,\sqrt{2}   & 686.1 \cdot 2     & 246.5 \cdot 2     & 115.3 \cdot 2     \\
      263.5 \,\sqrt{2}  & -43.1 \,\sqrt{2}  & -220.4 \,\sqrt{2} & 246.5 \cdot 2     & 327.9 \cdot 2     & -14.9 \cdot 2     \\
      -134.2 \,\sqrt{2} & 103.8 \,\sqrt{2}  & 30.4 \,\sqrt{2}   & 115.3 \cdot 2     & -14.9 \cdot 2     & 915.7 \cdot 2     \\
    \end{array}
  \right)
  \text{ GPa}^{2}.
\end{equation*}

\subsection{Elasticity tensor of Stahn--Müller--Bertram~\cite{SMB2020}}
\label{subsec:SMB2020}

The harmonic components of the elasticity tensor~\eqref{eq:E0Stahn} are:
\begin{equation*}
  \lambda_{0}=2.6913,
  \qquad
  \mu_{0}=3.9647,
\end{equation*}
\begin{equation*}
  \bd_{0}'=\left(
  \begin{array}{ccc}
      -1.0433 & 0.601  & -1.987  \\
      0.601   & 3.5667 & 2.9486  \\
      -1.987  & 2.9486 & -2.5233 \\
    \end{array}
  \right)
  ,
  \qquad
  \bv_{0}'=\left(
  \begin{array}{ccc}
      -1.165 & -0.382 & 1.1849 \\
      -0.382 & 0.67   & 0.3474 \\
      1.1849 & 0.3474 & 0.495  \\
    \end{array}
  \right),
\end{equation*}
and, by~\eqref{eq:H} (in Kelvin notation),
\begin{equation*}
  [\bH_{0}]=
  \left(
  \begin{array}{cccccc}
      -0.3069  & 1.0334    & -0.7266   & 2.40218   & 3.98271   & -2.47714  \\
      1.0334   & -0.9326   & -0.1009   & -1.5661   & -0.955301 & 2.82291   \\
      -0.7266  & -0.1009   & 0.8274    & -0.836083 & -3.02741  & -0.345917 \\
      2.40218  & -1.5661   & -0.836083 & -0.2018   & -0.4892   & -1.351    \\
      3.98271  & -0.955301 & -3.02741  & -0.4892   & -1.4532   & 3.3972    \\
      -2.47714 & 2.82291   & -0.345917 & -1.351    & 3.3972    & 2.0668    \\
    \end{array}
  \right),
\end{equation*}
giving
\begin{equation*}
  \bd_{20}'=\left(
  \begin{array}{ccc}
      19.749   & 11.0169  & -19.1923 \\
      11.0169  & -10.8286 & -3.795   \\
      -19.1923 & -3.795   & -8.9204  \\
    \end{array}
  \right).
\end{equation*}

Finally (in Kelvin notation):
\begin{equation*}
  [\bA_{0}]=
  \left(
  \begin{array}{cccccc}
      1.4587    & -0.4878   & -0.9709   & 0.608112  & 1.08668   & -0.662842 \\
      -0.4878   & 1.1331    & -0.6453   & -0.39103  & -0.115683 & 0.769049  \\
      -0.9709   & -0.6453   & 1.6162    & -0.217082 & -0.970999 & -0.106066 \\
      0.608112  & -0.39103  & -0.217082 & 1.8298    & 0.0152    & -0.4514   \\
      1.08668   & -0.115683 & -0.970999 & 0.0152    & 1.6372    & 0.803     \\
      -0.662842 & 0.769049  & -0.106066 & -0.4514   & 0.803     & 2.5748    \\
    \end{array}
  \right),
\end{equation*}
and
\begin{equation*}
  [\bB_{0}]=
  \left(
  \begin{array}{cccccc}
      -0.3187  & 0.315     & 0.0037    & 0.125158  & 2.30814  & -1.36401  \\
      0.315    & 2.1077    & -2.4227   & -0.149482 & -2.55845 & 0.0678823 \\
      0.0037   & -2.4227   & 2.4191    & 0.0243245 & 0.250316 & 1.29613   \\
      0.125158 & -0.149482 & 0.0243245 & 4.4958    & -1.472   & 2.1396    \\
      2.30814  & -2.55845  & 0.250316  & -1.472    & 0.1754   & 1.3154    \\
      -1.36401 & 0.0678823 & 1.29613   & 2.1396    & 1.3154   & 1.3706    \\
    \end{array}
  \right).
\end{equation*}

\subsection{Elasticity tensor of François~\cite{Fra1995} for Vosges sandstone}
\label{subsec:Fra1995}

The harmonic components of the elasticity tensor~\eqref{eq:E0Francois} are:
\begin{equation*}
  \lambda_{0}=1.327\;  \text{ GPa},
  \qquad
  \mu_{0}=5.393\;  \text{ GPa},
\end{equation*}
\begin{equation*}
  \bd_{0}'=\left(
  \begin{array}{ccc}
      -2.867 & 0.     & 0.5   \\
      0.     & -1.067 & 1.1   \\
      0.5    & 1.1    & 3.933 \\
    \end{array}
  \right)
  \text{ GPa},
  \qquad
  \bv_{0}'= \left(
  \begin{array}{ccc}
      0.1 & 1.1  & 0.8 \\
      1.1 & -0.5 & 0.2 \\
      0.8 & 0.2  & 0.4 \\
    \end{array}
  \right)
  \text{ GPa},
\end{equation*}
and, by~\eqref{eq:H} (in Kelvin notation),
\begin{equation*}
  [\bH_{0}]=\left(
  \begin{array}{cccccc}
      0.849             & -0.946            & 0.097             & 0.229 \,\sqrt{2}  & 0.5 \,\sqrt{2}  & -0.814 \,\sqrt{2} \\
      -0.946            & 1.477             & -0.531            & -0.014 \,\sqrt{2} & -0.2 \,\sqrt{2} & 0.086 \,\sqrt{2}  \\
      0.097             & -0.531            & 0.434             & -0.214 \,\sqrt{2} & -0.3 \,\sqrt{2} & 0.729 \,\sqrt{2}  \\
      0.229 \,\sqrt{2}  & -0.014 \,\sqrt{2} & -0.214 \,\sqrt{2} & -0.531 \cdot 2    & 0.729 \cdot 2   & -0.2 \cdot 2      \\
      0.5 \,\sqrt{2}    & -0.2 \,\sqrt{2}   & -0.3 \,\sqrt{2}   & 0.729 \cdot 2     & 0.097 \cdot 2   & 0.229 \cdot 2     \\
      -0.814 \,\sqrt{2} & 0.086 \,\sqrt{2}  & 0.729 \,\sqrt{2}  & -0.2 \cdot 2      & 0.229  \cdot 2  & -0.946 \cdot 2    \\
    \end{array}
  \right)
  \text{ GPa},
\end{equation*}
so that
\begin{equation*}
  \bd_{20}'=
  \left(
  \begin{array}{ccc}
      0.99  & 0.695 & 1.21  \\
      0.695 & 1.13  & -2.32 \\
      1.21  & -2.32 & -2.12 \\
    \end{array}
  \right)
  \text{ GPa}^{2}.
\end{equation*}

Finally (in Kelvin notation):
\begin{equation*}
  [\bA_{0}]=\left(
  \begin{array}{cccccc}
      0.213             & -0.13             & -0.083           & 0.059 \,\sqrt{2}  & -0.042 \,\sqrt{2} & -0.115 \,\sqrt{2} \\
      -0.13             & 0.241             & -0.111           & -0.039 \,\sqrt{2} & -0.014 \,\sqrt{2} & -0.008 \,\sqrt{2} \\
      -0.083            & -0.111            & 0.194            & -0.02 \,\sqrt{2}  & 0.056 \,\sqrt{2}  & 0.123 \,\sqrt{2}  \\
      0.059 \,\sqrt{2}  & -0.039 \,\sqrt{2} & -0.02 \,\sqrt{2} & 0.155 \cdot 2     & 0.128 \cdot 2     & -0.005 \cdot 2    \\
      -0.042 \,\sqrt{2} & -0.014 \,\sqrt{2} & 0.056 \,\sqrt{2} & 0.128 \cdot 2     & 0.182 \cdot 2     & 0.042 \cdot 2     \\
      -0.115 \,\sqrt{2} & -0.008 \,\sqrt{2} & 0.123 \,\sqrt{2} & -0.005 \cdot 2    & 0.042 \cdot 2     & 0.16 \cdot 2      \\
    \end{array}
  \right)
  \text{ GPa}^{2},
\end{equation*}
\begin{equation*}
  [\bB_{0}]=\left(
  \begin{array}{cccccc}
      0.124             & 0.061              & -0.185           & -0.15 \,\sqrt{2} & -0.096 \,\sqrt{2} & -0.146 \,\sqrt{2} \\
      0.061             & 0.139              & -0.2             & 0.065 \,\sqrt{2} & 0.094 \,\sqrt{2}  & -0.039 \,\sqrt{2} \\
      -0.185            & -0.2               & 0.385            & 0.085 \,\sqrt{2} & 0.002 \,\sqrt{2}  & 0.186 \,\sqrt{2}  \\
      -0.15 \,\sqrt{2}  & 0.065 \,\sqrt{2}   & 0.085 \,\sqrt{2} & 0.222 \cdot 2    & 0.081 \cdot 2     & -0.087 \cdot 2    \\
      -0.096 \,\sqrt{2} & 0.094 \,\sqrt{2}   & 0.002 \,\sqrt{2} & 0.081 \cdot 2    & 0.258 \cdot 2     & 0.198 \cdot 2     \\
      -0.146 \,\sqrt{2} & -0.039  \,\sqrt{2} & 0.186 \,\sqrt{2} & -0.087 \cdot 2   & 0.198 \cdot 2     & 0.017 \cdot 2     \\
    \end{array}
  \right)
  \text{ GPa}^{2}.
\end{equation*}

\section{Log-Euclidean upper bounds estimates}
\label{sec:Log}

For a given tensor $\bE_{0}$, once an elasticity tensor $\bE$ either cubic ($\bE\in \strata{\octa}$) or orthotropic
($\bE\in \strata{\DD_{2}}$) has been computed according to the symmetry group of a second-order tensor, say $\ba$, one can easily calculate the upper bounds estimates $\Delta_{\ba}(\bE_{0}, \strata{G})$ for any norm. For instance, since an elasticity tensor has to be positive definite, one can consider the Log-Euclidean norm~\cite{AFPA2005,MN2006},
\begin{equation*}
  \norm{\bE}_{L}:=\norm{\ln(\bE)}=\norm{\ln([\bE])}_{\RR^{6}},
\end{equation*}
which has the property of invariance by inversion, meaning that the norm for the compliance is the same as the norm for the stiffness  ($\norm{\bE^{-1}}_{L}=\norm{\bE}_{L}$). For this norm, the upper bounds estimates of the distance $d(\bE_{0}, \strata{G})=\min_{\strata{G}}\norm{\bE_{0}-\bE}_{L}$ to the symmetry stratum $\strata{G}$ can then be expressed as
\begin{equation*}
  \Delta_{\ba}(\bE_{0}, \strata{G}):=\norm{\bE_{0}-\bE}_{L}=\norm{\ln(\bE_{0})-\ln (\bE)},
\end{equation*}

\begin{rem}
  This is $\bE$ that is enforced to be either cubic (by formulas ~\eqref{eq:Ca}--\eqref{eq:DeltaNous1}) or orthotropic (by formulas~\eqref{eq:UpperOrtho}--\eqref{eq:alphabetagamma}), not $\ln(\bE)$~\cite{MN2006}. One does not perform the harmonic decomposition of neither $\ln(\bE_{0})$ nor $\ln(\bE)$ and formula~\eqref{eq:DeltaNous2} does not apply anymore.
\end{rem}

We provide in the following tables the comparisons (for the single crystal superalloy~\eqref{eq:E0} and the Vosges sandstone~\eqref{eq:E0Francois}) of the relative upper bounds estimates obtained with the Euclidean norm,
\begin{equation*}
  \frac{\Delta_{\ba}(\bE_{0}, \strata{G})}{\norm{\bE_{0}}}=\frac{\norm{\bE_{0}-\bE}}{\norm{\bE}},
\end{equation*}
and with the Log-Euclidean norm,
\begin{equation*}
  \frac{\Delta_{\ba}(\bE_{0}, \strata{G})}{\norm{\bE_{0}}_{L}}=\frac{\norm{\bE_{0}-\bE}_{L}}{\norm{\bE}_{L}}=\frac{\norm{\ln(\bE_{0})-\ln (\bE)}}{\norm{\ln(\bE)}}.
\end{equation*}
The second-order tensor $\ba$ is taken in the list $\set{\bt_{0}', \bd_{20}', \ba', \bb'}$ computed in sections \ref{sec:Upper bound-Cubic} and \ref{sec:Upper bound-Ortho}. The minimum minimorum is also given.

We do not provide comparisons for the academic tensor~\eqref{eq:E0Stahn} since it is not positive definite.
Note that the order of magnitude of the relative distance strongly depends on the choice of a norm.

\begin{table}[h]
  \begin{center}
    \setlength{\arraycolsep}{1pt}
    \begin{tabular}{ccccc|c}
      \toprule
                                       & $\Delta_{\bt_{0}}$    & $ \Delta_{\bd_{20}}$ &
      $ \Delta_{\ba'}$                  &
      $ \Delta_{\bb'}$                  & $\Delta_{\text{opt}}$                                                   \\
      \midrule
      Relative Euclidean estimate:     & 0.3388                & 0.3344               & 0.1610 & 0.1371 & 0.1371 \\
      Relative Log-Euclidean estimate: & 0.1365                & 0.1353               & 0.0616 & 0.0516 & 0.0516 \\
      \bottomrule
    \end{tabular}
    \caption{Comparison of cubic upper bounds estimates for Ni-based single crystal superalloy.}
  \end{center}
\end{table}

\begin{table}[h]
  \begin{center}
    \setlength{\arraycolsep}{1pt}
    \begin{tabular}{ccccc|c}
      \toprule
                                       & $\Delta_{\bt_{0}}$    & $ \Delta_{\bd_{20}}$ &
      $ \Delta_{\ba'}$                  &
      $ \Delta_{\bb'}$                  & $\Delta_{\text{opt}}$                                                   \\
      \midrule
      Relative Euclidean estimate:     & 0.2660                & 0.2664               & 0.2674 & 0.2596 & 0.2596 \\
      Relative Log-Euclidean estimate: & 0.1261                & 0.1276               & 0.1274 & 0.1256 & 0.1256 \\
      \bottomrule
    \end{tabular}
    \caption{Comparison of cubic upper bounds estimates for Vosges sandstone.}
  \end{center}
\end{table}

\begin{table}[h]
  \begin{center}
    \setlength{\arraycolsep}{1pt}
    \begin{tabular}{ccccc|c}
      \toprule
                                       & $\Delta_{\bt_{0}'}$   & $ \Delta_{\bd_{20}'}$ &
      $ \Delta_{\ba'}$                 &
      $ \Delta_{\bb'}$                 & $\Delta_{\text{opt}}$                                                    \\
      \midrule
      Relative Euclidean estimate:     & 0.3029                & 0.2943                & 0.1539 & 0.1266 & 0.1266 \\
      Relative Log-Euclidean estimate: & 0.1221                & 0.1160                & 0.0529 & 0.0392 & 0.0392 \\
      \bottomrule
    \end{tabular}
    \caption{Comparison of orthotropic upper bounds estimates for Ni-based single crystal superalloy.}
  \end{center}
\end{table}

\begin{table}[h]
  \begin{center}
    \setlength{\arraycolsep}{1pt}
    \begin{tabular}{ccccc|c}
      \toprule
                                       & $\Delta_{\bt_{0}'}$   & $ \Delta_{\bd_{20}'}$ &
      $ \Delta_{\ba'}$                 &
      $ \Delta_{\bb'}$                 & $\Delta_{\text{opt}}$                                                      \\
      \midrule
      Relative Euclidean estimate:     &
      0.14907                          & 0.15531               & 0.16453               & 0.15530 & 0.14907          \\
      Relative Log-Euclidean estimate: & 0.0685                & 0.0728                & 0.0749  & 0.0786  & 0.0685 \\
      \bottomrule
    \end{tabular}
    \caption{Comparison of orthotropic upper bounds estimates for Vosges sandstone.}
  \end{center}
\end{table}



\end{document}